\documentclass[aps,prl,twocolumn,superscriptaddress,showkeys,amsmath,amssymb,floatfix]{revtex4-2}

\usepackage{graphicx,bm,booktabs}
\usepackage[percent]{overpic}
\usepackage{dcolumn}
\usepackage{hyperref}
\usepackage{xcolor}
\usepackage[normalem]{ulem}
\usepackage{tikz}
\usetikzlibrary{decorations.pathmorphing, decorations.markings, arrows.meta}
\usepackage{physics}

\hypersetup{
	colorlinks=true,
	linkcolor=blue,
	filecolor=magenta,
	urlcolor=blue,
	citecolor=blue,
}

\graphicspath{{figures/}}

\definecolor{lightblue}{RGB}{174,215,236}
\definecolor{layeredge}{RGB}{62,87,101}
\definecolor{currentblue}{RGB}{0,105,180}
\definecolor{phasecolor}{RGB}{125,45,160}

\newcommand{\kc}{k_0^{c}}
\newcommand{\phic}{\phi_{\rm c}}
\newcommand{\phir}{\phi_{\rm r}}
\newcommand{\rhosol}{\rho_{\rm sol}}

\begin{document}
	
	\title{Interlayer Phase Locking and Sliding Select Topological Melting Channels in Superconducting Bilayers}
	
	\author{Yicheng Zhong}
	\affiliation{Institute of Physics, Chinese Academy of Sciences, Beijing 100190, China}
	\affiliation{School of Physical Sciences, University of Chinese Academy of Sciences, Beijing 101408, China}
	
	\author{Yi Zhou}
	\email{yizhou@iphy.ac.cn}
	\affiliation{Institute of Physics, Chinese Academy of Sciences, Beijing 100190, China}
	
	\date{\today}
	
	\begin{abstract}
		Topological transitions in coupled superconducting layers are intertwined: an elementary vortex in one layer carries both common and relative phase winding. We study a Josephson-coupled bilayer in an in-plane magnetic field using renormalization-group analysis and Monte Carlo simulations of the compact layer phases. The field drives Pokrovsky--Talapov (PT) entry of interlayer solitons, and free-energy calculations in fixed soliton sectors show PT square-root scaling over the sampled densities at finite temperature. Interlayer phase locking and sliding then select distinct Berezinskii--Kosterlitz--Thouless (BKT)-like melting channels. The locked commensurate phase melts through same-vorticity layer pairs when the geometric-mean common helicity modulus $\Upsilon_{+,g}$ reaches $T/(2\pi)$, with layer correlation exponent $\eta_l=1/4$. A stable sliding incommensurate phase melts through elementary layer vortices on a line of BKT fixed points, where $\Upsilon_{+,g}+\Upsilon_{-,g}=2T/\pi$ and $\eta_l$ is set by the ratio of the relative and common moduli. When the relative helicity modulus is identified with the relative-phase stiffness, our simulations agree with this prediction, and the common superfluid stiffness at incommensurate melting is nearly twice its value at commensurate melting. Near soliton entry, however, the dilute sliding state is unstable to opposite-vorticity layer pairs. Our results connect PT soliton entry with field-tunable BKT-like melting and identify the pair-stability condition that links them.
	\end{abstract}

	
	\maketitle

	\section{Introduction}
		In two dimensions (2D), superfluids and superconductors melt through the unbinding of topological defects, the Berezinskii--Kosterlitz--Thouless (BKT) mechanism~\cite{Berezinskii1971,KosterlitzThouless1973,NelsonKosterlitz1977,Beasley1979,HalperinNelson1979,Minnhagen1987}. When two condensates are coupled, as in layered superconductors and coupled superfluids~\cite{Benfatto2007,Mathey2008}, which defects unbind depends on whether their relative phase is locked. An in-plane magnetic field $H$ makes this tunable in a Josephson-coupled superconducting bilayer, where it can drive the relative phase from locked to sliding~\cite{QiuZhou2022}, as in double-layer quantum Hall systems~\cite{KYang94,KYang96}.

	\begin{figure*}[tb]
		\centering
		\begin{overpic}[width=0.47\textwidth]{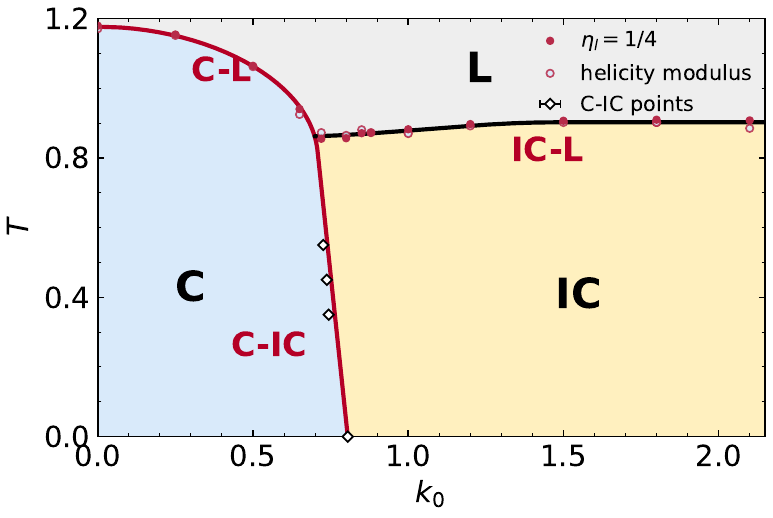}
			\put(-1,68){\bfseries (a)}
		\end{overpic}
		\hfill
		\begin{overpic}[width=0.47\textwidth]{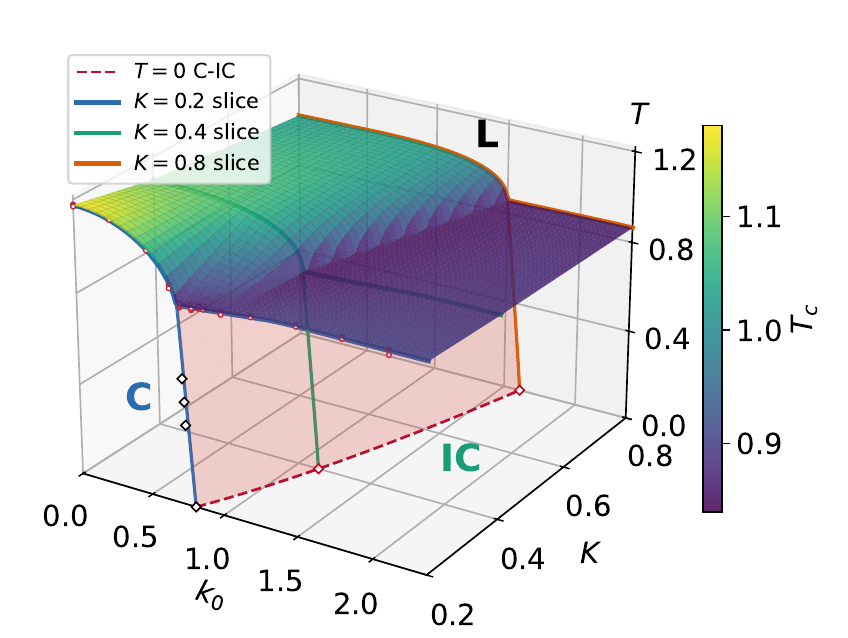}
			\put(-1,68){\bfseries (b)}
		\end{overpic}
		\caption{Phase diagrams of the compact layer-phase model. Here $k_0$ is the field-induced momentum mismatch and $K$ the phase-only Josephson coupling; C, IC, and L denote the commensurate, incommensurate, and thermal-liquid phases. (a) Cut at $K=0.2$. Diamonds mark the intercepts $\kc(T)$ of the PT fits in Fig.~\ref{fig:ci} and, at $T=0$, the continuum threshold of Eq.~\eqref{eq:kc}. Open circles mark C--L estimates from Eq.~\eqref{eq:lockedcriterion} and IC helicity-modulus crossings from Eq.~\eqref{eq:singlecriterion}; filled circles mark crossings of the layer correlation exponent at $\eta_l=1/4$, shown on the IC--L branch for comparison. Curves are guides to the eye: the straight C--IC line follows the trend from $T=0$ rather than interpolating the finite-temperature intercepts, and its junction with the melting curves is extrapolated. (b) Schematic $(k_0,T,K)$ phase diagram. The $K=0.2$ slice reproduces (a); the $K=0.4$ and $0.8$ curves are schematic guides, and the surfaces interpolate between slices. The dashed line marks the $T=0$ threshold of Eq.~\eqref{eq:kc}.}
		\label{fig:phase_diagram}
	\end{figure*}

\begin{figure*}[tb]
	\centering
	\resizebox{\linewidth}{!}{\begin{tikzpicture}[font=\sffamily,>=Latex,layer/.style={draw=layeredge,line width=0.55pt},current/.style={-{Latex[length=2.3mm,width=1.7mm]},line width=0.95pt,draw=currentblue},currentthin/.style={-{Latex[length=1.8mm,width=1.35mm]},line width=0.80pt,draw=currentblue},josephson/.style={-{Latex[length=1.6mm,width=1.2mm]},line width=0.80pt,draw=currentblue},vortexbase/.style={line width=1.25pt,line cap=round,postaction={decorate},decoration={markings,mark=at position 0.56 with {\arrow{Latex[length=1.8mm,width=1.35mm]}}}},vortexred/.style={vortexbase,draw=red!85},vortexyellow/.style={vortexbase,draw=orange!85!yellow},cellsep/.style={draw=black!45,densely dashed,line width=0.55pt},density/.style={draw=black!75,line width=0.65pt},thinblack/.style={draw=black!75,line width=0.45pt},field/.style={-{Latex[length=2.2mm,width=1.6mm]},line width=1.0pt,draw=red!80},phase/.style={draw=phasecolor,line width=1.3pt},phaseaxis/.style={draw=black!60,line width=0.45pt},phaseguide/.style={draw=black!25,densely dotted,line width=0.45pt},declare function={kink(\u)=0.5*(1+tanh(max(min(\u,6),-6)));}]
		\path[fill=white] (-0.35,1.98) rectangle (17.75,5.42); \path[use as bounding box] (-0.35,1.98) rectangle (17.75,5.42);
		\def\xL{0.40}\def\xR{5.10}\def\yT{4.45}\def\yB{3.25}\def\yM{3.85}\def\layerH{0.13}
		\def\yP{2.21}\def\hP{0.33}
		\begin{scope}[shift={(0,0)}]
			\node[anchor=west,font=\rmfamily\small] at (-0.05,5.20) {\textbf{(a)}\enspace Bilayer geometry};
			\def\dX{0.95}\def\dY{0.62}\def\th{0.14}
			\def\slab#1#2{
				\path[layer,fill=lightblue] (0.55,#1) -- (3.95,#1) -- (3.95+\dX,#1+\dY) -- (0.55+\dX,#1+\dY) -- cycle;
				\path[layer,fill=lightblue!70!layeredge] (0.55,#1) -- (3.95,#1) -- (3.95,#1-\th) -- (0.55,#1-\th) -- cycle;
				\path[layer,fill=lightblue!45!layeredge] (3.95,#1) -- (3.95+\dX,#1+\dY) -- (3.95+\dX,#1+\dY-\th) -- (3.95,#1-\th) -- cycle;
				\node[anchor=west,font=\sffamily\small] at (3.95+\dX+0.05,#1+0.5*\dY-0.05) {#2};}
			\slab{2.75}{$l=2$}
			\foreach \gx in {3.20,3.38,3.56}{\draw[thinblack,decorate,decoration={snake,amplitude=0.035cm,segment length=0.12cm}] (\gx+0.48,2.75+0.31) -- (\gx+0.48,4.05-\th);}
			\node[anchor=west,font=\sffamily\small] at (4.12,3.50) {$g$};
			\foreach \fx in {0.95,1.85,2.75}{\draw[field] (\fx,3.15) -- ++(0.85*\dX,0.85*\dY);}
			\node[red!85,anchor=north,font=\sffamily\small] at (2.35,3.12) {$\mathbf H=H\hat y$};
			\slab{4.05}{$l=1$}
			\draw[thinblack,<->] (0.38,2.75) -- (0.38,4.05-\th); \node[anchor=east] at (0.34,3.33) {$a$};
			\coordinate (frameO) at (4.32,2.10);
			\draw[thinblack,->] (frameO) -- ++(0.45,0) node[anchor=west,font=\sffamily\small,inner sep=1pt] {$x$};
			\draw[thinblack,->] (frameO) -- ++(0.34,0.22) node[anchor=south west,font=\sffamily\small,inner sep=1pt] {$y$};
			\draw[thinblack,->] (frameO) -- ++(0,0.42) node[anchor=south,font=\sffamily\small,inner sep=1pt] {$z$};
		\end{scope}
		\begin{scope}[shift={(5.95,0)}]
			\node[anchor=west,font=\rmfamily\small] at (-0.05,5.20) {\textbf{(b)}\enspace Fulde--Ferrell state};
			\path[layer,fill=currentblue!32!white] (\xL,\yT-\layerH) rectangle (\xR,\yT+\layerH); \path[layer,fill=currentblue!32!white] (\xL,\yB-\layerH) rectangle (\xR,\yB+\layerH);
			\foreach \x in {0.70,1.55,2.40,3.25,4.10}{\draw[current] (\x+0.52,\yT) -- ++(-0.52,0); \draw[current] (\x,\yB) -- ++(0.52,0);}
			\node[anchor=center,font=\sffamily\small,text=currentblue] at (2.75,\yM) {$J_T=0$};
			\foreach \k in {1,2}{\draw[phaseguide] (\xL,\yP+\k*\hP) -- (\xR,\yP+\k*\hP);}
			\draw[phaseaxis] (\xL,\yP) -- (\xR,\yP);
			\foreach \k/\lab in {0/0,1/2\pi,2/4\pi}{\node[anchor=west,font=\sffamily\scriptsize] at (\xR+0.04,\yP+\k*\hP) {$\lab$};}
			\node[anchor=east,font=\sffamily\small,inner sep=1pt] at (\xL-0.04,\yP+\hP) {$\theta_-$};
			\draw[phase] (\xL,\yP) -- (\xR,\yP);
			\node[anchor=south,font=\sffamily\small,text=phasecolor,inner sep=1pt] at (2.75,\yP+0.05) {locked, $\theta_-=0$};
		\end{scope}
		\begin{scope}[shift={(11.90,0)}]
			\node[anchor=west,font=\rmfamily\small] at (-0.05,5.20) {\textbf{(c)}\enspace Bloch SC state}; \def\xcA{0.62}\def\xcB{2.75}\def\xcC{4.88}
			\shade[left color=currentblue!14!white,right color=currentblue!50!white,draw=layeredge,line width=0.45pt] (4.17,5.13) rectangle (4.77,5.27);
			\node[anchor=east,font=\sffamily\scriptsize,inner sep=1pt] at (4.14,5.20) {low}; \node[anchor=west,font=\sffamily\scriptsize,inner sep=1pt] at (4.80,5.20) {high};
			\node[anchor=east,font=\sffamily\small,inner sep=1pt] at (3.66,5.20) {$n_{s,l}$};
			\foreach \i in {0,...,93}{\pgfmathsetmacro{\xs}{\xL+\i*0.05}\pgfmathsetmacro{\pct}{int(14+36*(0.5-0.5*cos(360*(\xs+0.025-\xcA)/2.13)))}
				\fill[currentblue!\pct!white] (\xs,\yT-\layerH) rectangle (\xs+0.051,\yT+\layerH); \fill[currentblue!\pct!white] (\xs,\yB-\layerH) rectangle (\xs+0.051,\yB+\layerH);}
			\path[layer] (\xL,\yT-\layerH) rectangle (\xR,\yT+\layerH); \path[layer] (\xL,\yB-\layerH) rectangle (\xR,\yB+\layerH);
			\foreach \x in {\xcA,\xcB,\xcC}{\draw[cellsep] (\x,\yB-0.18) -- (\x,\yT+0.18);}
			\foreach \xs/\xe in {0.68/0.94,1.08/1.52,2.33/1.89,2.68/2.42,2.81/3.07,3.21/3.65,4.46/4.02,4.81/4.55}{\draw[currentthin] (\xs,\yT) -- (\xe,\yT); \draw[currentthin] (\xe,\yB) -- (\xs,\yB);}
			\foreach \x/\dy in {0.50/0.46,0.74/0.46,1.58/-0.46,1.82/-0.46,2.63/0.46,2.87/0.46,3.71/-0.46,3.95/-0.46,4.76/0.46,5.00/0.46}{\draw[josephson] (\x,\yM-0.5*\dy) -- (\x,\yM+0.5*\dy);}
			\foreach \x in {0.62,1.70,2.75,3.83,4.88}{\fill[currentblue] (\x,\yT) circle (0.025); \fill[currentblue] (\x,\yB) circle (0.025);}
			\foreach \x in {1.16,2.225,3.29,4.355}{\fill[currentblue] (\x,\yM) circle (0.020);}
			\def\loopUp#1#2#3{\draw[#1] (#2,\yM-0.34) .. controls (#3,\yM-0.23) and (#3,\yM+0.23) .. (#2,\yM+0.34);}
			\def\loopDown#1#2#3{\draw[#1] (#2,\yM+0.34) .. controls (#3,\yM+0.23) and (#3,\yM-0.23) .. (#2,\yM-0.34);}
			\loopUp{vortexyellow}{1.04}{0.86}\loopDown{vortexyellow}{1.28}{1.46}\loopDown{vortexred}{2.105}{1.925}\loopUp{vortexred}{2.345}{2.525}
			\loopUp{vortexyellow}{3.17}{2.99}\loopDown{vortexyellow}{3.41}{3.59}\loopDown{vortexred}{4.235}{4.055}\loopUp{vortexred}{4.475}{4.655}
			\foreach \k in {1,2}{\draw[phaseguide] (\xL,\yP+\k*\hP) -- (\xR,\yP+\k*\hP);}
			\draw[phaseaxis] (\xL,\yP) -- (\xR,\yP);
			\foreach \k/\lab in {0/0,1/2\pi,2/4\pi}{\node[anchor=west,font=\sffamily\scriptsize] at (\xR+0.04,\yP+\k*\hP) {$\lab$};}
			\node[anchor=east,font=\sffamily\small,inner sep=1pt] at (\xL-0.04,\yP+\hP) {$\theta_-$};
			\foreach \xk in {1.16,3.29}{\draw[phaseguide] (\xk,\yP) -- (\xk,\yB-0.18);}
			\draw[phase,domain=\xL:\xR,samples=160,smooth] plot ({\x},{\yP+\hP*(kink((\x-1.16)/0.20)+kink((\x-3.29)/0.20))});
			\node[anchor=west,font=\sffamily\small,text=phasecolor,inner sep=1pt] at (1.40,\yP+0.42*\hP) {$2\pi$ soliton};
		\end{scope}
	\end{tikzpicture}}
	\caption{Bilayer geometry, currents, and relative phase. (a)~Layers $l=1,2$, separated by $a$ and coupled by Josephson tunneling $g$, in an in-plane field $\mathbf H=H\hat y$. (b)~Fulde--Ferrell state: uniform superfluid density $n_{s,l}$, counterpropagating screening currents, vanishing interlayer current ($J_T=0$), and locked relative phase $\theta_-=0$. (c)~Bloch SC state: modulated $n_{s,l}$ and a lattice of $2\pi$ solitons in $\theta_-$. Shading indicates $n_{s,l}$; blue arrows show intralayer and Josephson currents, with lengths indicating magnitudes; colored loops and dashed lines mark vortex--antivortex unit cells; lower panels sketch $\theta_-(x)$.}
	\label{fig:model}
\end{figure*}

		The same bilayer is a minimal, tunable realization of finite-momentum superconductivity~\cite{Bulaevskii1973,Bulaevskii1975,LawrenceDoniach,BulaevskiiLedvijKogan1992,BuzdinKachkachi1997,Agterberg2003,Barzykin2002,Aoyama2012,Dimitrova2003,QiuZhou2022,Nag2025}, in which orbital coupling shifts the pair momentum oppositely in the two layers. Finite-momentum pairing is a recurring theme in 2D systems: it has been studied in correlated moir\'e bilayers~\cite{Cao2020,Wang2022,Xia2025} and in Ising superconductors in transition-metal dichalcogenides (TMDs)~\cite{Lu2015,Saito2016a,Xi2016,Hamill2021,Wan2023}, in-plane-field-induced pairing has been proposed in bilayer TMDs~\cite{bilayerTMD}, and segmented-Fermi-surface measurements show signatures of finite Cooper-pair momentum~\cite{Zheng21}. These states belong to the broader Fulde--Ferrell--Larkin--Ovchinnikov (FFLO) and pair-density-wave family, which includes Josephson-coupled FFLO systems~\cite{FuldeFerrell1964,LarkinOvchinnikov1965,Yang2000,CasalbuoniNardulli2004,ARCMP20}.
	
		At zero temperature, the field selects between two states~\cite{QiuZhou2022}. The Fulde--Ferrell (FF) state has uniform superfluid density and counterpropagating layer supercurrents. The Bloch superconducting (Bloch SC) state has a modulated superfluid density and a lattice of interlayer Josephson vortex--antivortex pairs, an alternating texture distinct from conventional Josephson vortex structures~\cite{Koshelev2013,Berdiyorov2018,Curran2018}. Increasing $H$ drives a commensurate--incommensurate (C--IC) transition from FF to Bloch SC, a 2D Pokrovsky--Talapov (PT) transition~\cite{PokrovskyTalapov1979,PokrovskyTalapov1980,Coppersmith1981,Coppersmith1982,Bak1982}. Self-consistent Bogoliubov--de Gennes (BdG) calculations support this picture, finding a stripe state with localized relative-phase vortices, gap-amplitude oscillations, and alternating supercurrents~\cite{Nag2025}.
	
	In the bilayer, Josephson coupling favors a locked relative phase, as in the FF state, whereas the in-plane field drives the relative phase to wind; their competition produces the soliton lattice of the Bloch SC state. The vortex-free theory separates into common and relative sectors, but because the physical layer phases are compact, each elementary layer vortex carries both common and relative winding. Soliton entry and vortex unbinding are therefore coupled. The central question raised by this coupling is how interlayer phase locking and sliding reorganize thermal melting.
	
	Figure~\ref{fig:phase_diagram} summarizes our numerical estimates and the resulting phase structure. We denote the locked and sliding bilayer superfluids, the finite-temperature continuations of the FF and Bloch SC states, by C/FF and IC/Bloch SC, and the thermal liquid by L; C--L and IC--L denote the corresponding melting transitions. Fixed-soliton-sector free-energy crossings show PT scaling of soliton entry at finite temperature. Renormalization-group (RG) analysis shows that same-vorticity layer pairs melt the locked phase, whereas elementary layer vortices melt a stable sliding phase, in which both pair-vortex channels are irrelevant. Near soliton entry, opposite-vorticity pairs destabilize the dilute sliding state. Helicity-modulus crossings and correlation exponents support the two melting channels on representative cuts.

	\section{Model and methods}
	\subsection{Effective theory}
	We consider two identical superconducting films of thickness $d$ much smaller than the coherence length $\xi$ and penetration depth $\lambda$, separated by a distance $a$, weakly coupled by Josephson tunneling, and placed in an in-plane field $\mathbf{H}=H\hat{y}$ [Fig.~\ref{fig:model}(a)]. Let $\psi_l$ and $\mathbf A_l$ denote the order parameter and vector potential in layer $l$. For Cooper-pair charge $e^*=2e$, the 2D Ginzburg--Landau free-energy density is~\cite{LawrenceDoniach,QiuZhou2022}
	\begin{equation}
	\begin{split}
		f &= f_n + \sum_{l=1,2}\!\bigg[ \alpha|\psi_l|^2 + \tfrac{\beta}{2}|\psi_l|^4 \\
		& \quad + \frac{1}{2m^*}\Big|\!\Big(\frac{\hbar}{i}\nabla - \frac{e^*}{c}\mathbf{A}_l\!\Big)\psi_l\Big|^2\bigg] + g\bigl(\psi_1^*\psi_2 + \mathrm{c.c.}\bigr).
	\end{split}
	\label{eq:GL}
	\end{equation}
	Here $f_n$ is the normal-state free-energy density, $\alpha$ and $\beta$ are the intralayer Ginzburg--Landau coefficients, $m^*$ is the Cooper-pair effective mass, and $g$ is the Josephson coupling; the film thickness is absorbed into these coefficients.
	In the Landau gauge $\mathbf A=Hz\,\hat{x}$, with the layers at $z=s_l a/2$, $s_1=+1$, and $s_2=-1$, the layer vector potentials are $\mathbf A_l=s_l(Ha/2)\hat{x}$. For local tunneling between vertically aligned points, $\mathbf A\cdot d\mathbf s=0$ along the tunneling path, so the Josephson term acquires no gauge phase. Instead, the field shifts the intralayer phase gradients oppositely, by $s_lk_0\hat{x}/2$ in layer $l$~\cite{QiuZhou2022}, with
	\begin{equation}
		k_0 \;=\; 2\pi H a / \Phi_0 ,
		\label{eq:k0}
	\end{equation}
	where $\Phi_0=hc/e^*$ is the flux quantum.

		Because phase fluctuations govern the transitions, we set $\psi_l=\Psi e^{i\theta_l}$ with constant amplitude $\Psi$. This phase-only approximation captures the solitons and currents in Fig.~\ref{fig:model}(c) but omits the accompanying superfluid-density modulation. In terms of the common and relative phases $\theta_\pm=\theta_1\pm\theta_2$, the free energy becomes
	\begin{equation}
	\begin{split}
		\mathcal{F}
		&= \int d^2\mathbf{r}\,
		\bigg[
		\frac{\rho_s}{4}(\nabla\theta_+)^2
		+\frac{\rho_s}{4}(\nabla\theta_- - k_0\hat{x})^2
		-K\cos\theta_-
		\bigg].
	\end{split}
	\label{eq:Feff}
	\end{equation}
		Here $\rho_s=\hbar^2\Psi^2/m^*$ is the bare single-layer phase stiffness, and $K=2|g|\Psi^2$ is the phase-only Josephson coupling, with the Josephson-energy minimum chosen at $\theta_-=0$. In the vortex-free sector, the two phases decouple: the field-induced tilt $-\rho_s k_0\partial_x\theta_-/2$ and the lock-in term $-K\cos\theta_-$ act only on $\theta_-$. Thus $\theta_+$ is Gaussian, whereas $\theta_-$ realizes a 2D PT model with commensurability $p=1$~\cite{PokrovskyTalapov1979,PokrovskyTalapov1980,Coppersmith1981,Coppersmith1982,Horovitz1983,HaldaneBakBohr1983,KYang94,KYang96}. On the incommensurate side, $\theta_-$ forms the soliton lattice that produces the Bloch SC currents in Fig.~\ref{fig:model}(c).
	
		The physical layer phases are compact, $\theta_l\equiv\theta_l+2\pi$, so $\theta_\pm$ are not independently compact. Layer windings $(w_1,w_2)\in\mathbb Z^2$ give common and relative windings $w_\pm=w_1\pm w_2$. An elementary vortex in either layer thus carries $(w_+,w_-)=(1,\pm1)$, a same-vorticity layer pair carries $(2,0)$, and an opposite-vorticity pair carries $(0,2)$. Because elementary layer vortices carry both windings, their role in melting depends on whether the relative phase is locked or sliding.

	\subsection{Monte Carlo simulations}
		We simulate the compact layer-phase model directly in the physical phases $\theta_1$ and $\theta_2$ on square lattices of unit spacing, in units with $\rho_s=1$, at $K=0.2$; the lattice action, sampling protocol, boundary conditions, and error analysis are given in Appendix~\ref{app:mc}. The thermal cuts, at fixed $k_0$ and varying $T$, use linear size $L=100$, and the fixed-sector C--IC analysis combines $L=64$, 100, and 128.

	\section{Results}
	\subsection{Commensurate--incommensurate transition}
		At zero temperature, the ground state contains no free layer vortices, so the vortex-free relative phase sets the C--IC threshold. A straight $2\pi$ sine-Gordon wall costs line tension $8\sqrt{\rho_s K/2}$ but gains tilt energy $\pi\rho_s k_0$ per unit length. Walls therefore enter at
	\begin{equation}
		\kc=\frac{4}{\pi}\sqrt{\frac{2K}{\rho_s}}.
		\label{eq:kc}
	\end{equation}
		Fixed-sector lattice minimization at $K=0.8$ reproduces Eq.~\eqref{eq:kc} to within 0.5\% and confirms the inverse-logarithmic onset expected for a dilute soliton lattice (Appendix~\ref{app:zeroT}).

		At finite temperature, thermal wandering of the walls changes the onset of the soliton density $\rho_{\rm sol}$, the number of solitons per unit length along $x$, to the PT form $\rho_{\rm sol}\propto[k_0-\kc(T)]^{1/2}$, with a finite-temperature threshold $\kc(T)$, which we test below with fixed-sector free-energy crossings. In the asymptotic dilute limit, however, opposite-vorticity layer pairs destabilize the sliding state, as the RG analysis below shows, so a stable sliding phase requires a sufficiently dense soliton lattice.
	
	\subsection{BKT-like thermal melting}
		Layer vortices, not Josephson solitons, drive thermal melting. We set $k_{\rm B}=1$. The common ($+$) and relative ($-$) helicity moduli $\Upsilon_{\pm,x}$ and $\Upsilon_{\pm,y}$ are defined by uniform phase offsets in the intralayer couplings along $x$ and $y$ and normalized so that their bare harmonic values are $\rho_s/2$ (Appendix~\ref{app:mc}). Because vortex scaling with diagonal anisotropy depends on geometric means of the directional stiffnesses, we use $\Upsilon_{\pm,g}=\sqrt{\Upsilon_{\pm,x}\Upsilon_{\pm,y}}$. In the IC phase, the relative-phase stiffness describes slow compression and tilt of the soliton lattice and need not equal the relative helicity modulus.

		In the C phase, Josephson locking confines every defect with relative winding, leaving the same-vorticity layer pair $(w_1,w_2)=(1,1)$ as the active BKT defect. Its common winding $w_+=2$ sets the melting criterion
	\begin{equation}
		\Upsilon_{+,g}=T/(2\pi).
		\label{eq:lockedcriterion}
	\end{equation}
		At this boundary the physical-layer correlation exponent $\eta_l$, defined by $\langle e^{i[\theta_l(\mathbf r)-\theta_l(0)]}\rangle\propto r^{-\eta_l}$, equals $1/4$, which provides an independent correlation test.
	
		In the sliding IC phase, both phase modes are massless. When both pair-vortex channels are irrelevant, elementary layer vortices, each carrying $(w_+,w_-)=(1,\pm1)$, control melting. Their marginality requires $\mathcal K_{\rm c}+\mathcal K_g=2/\pi$, where $\mathcal K_{\rm c}$ and $\mathcal K_g$ are the dimensionless common- and relative-phase stiffnesses defined below. With $\mathcal K_{\rm c}=\Upsilon_{+,g}/T$ and $\mathcal K_g=\Upsilon_{-,g}/T$, this becomes
	\begin{equation}
		\Upsilon_{+,g}+\Upsilon_{-,g}=2T/\pi.
		\label{eq:singlecriterion}
	\end{equation}
		The identification $\mathcal K_g=\Upsilon_{-,g}/T$ is an assumption that the present data do not test independently; we adopt it for all IC comparisons and use Eq.~\eqref{eq:singlecriterion} to locate IC melting, as Eq.~\eqref{eq:lockedcriterion} locates C--L melting. At criticality, $r=\Upsilon_{-,g}/\Upsilon_{+,g}$ sets the critical layer exponent $\eta_l^*=(1+r)^2/(16r)\geq1/4$, with equality only at $r=1$.
	
	\subsection{Renormalization-group analysis}
		We derive these melting criteria from an RG treatment that follows the fixed-density formulation of Horovitz \emph{et al.}~\cite{Horovitz1983}: we write $\theta_-=\varphi+Qx$, with $Q=2\pi\rho_{\rm sol}$, and coarse grain the periodic field $\varphi$ at fixed $\rho_{\rm sol}$. Solitons make the relative sector anisotropic, and vortex screening can transmit anisotropy to both sectors. We use the geometric means $\mathcal K_{\rm c}$ and $\mathcal K_g$ of the directional common and relative stiffnesses, whose bare values are $\rho_s/(2T)$ (Appendix~\ref{app:rg}).
	
		We introduce fugacities $y_J$, $y_1$, $y_s$, and $y_a$ for Josephson lock-in, elementary layer vortices, same-vorticity pairs, and opposite-vorticity pairs, as in related compact-phase RG schemes~\cite{Benfatto2007,Mathey2008}, using positive vortex activities and a circular real-space separation shell (Appendix~\ref{app:rg}). At the microscopic cutoff $a_0$, $y_J=Ka_0^2/T$, and the vortex fugacities start from Boltzmann factors of their core energies. With $\ell$ the logarithmic scale of the running cutoff $a_0e^\ell$, the fugacities flow to second order as
	\begin{align}
		\frac{dy_J}{d\ell}&=\left(2-\frac{1}{4\pi \mathcal K_g}\right)y_J,\nonumber \\
		\frac{dy_1}{d\ell}&=\left[2-\pi(\mathcal K_{\rm c}+\mathcal K_g)+O(y_s,y_a)\right]y_1,\nonumber \\
		\frac{dy_s}{d\ell}&=\left(2-4\pi \mathcal K_{\rm c}\right)y_s+c_s y_1^2,\nonumber \\
		\frac{dy_a}{d\ell}&=\left(2-4\pi \mathcal K_g\right)y_a+c_a y_1^2,
	\label{eq:compactRG}
	\end{align}
	where the pair-vortex feedback on $y_1$ and the fusion sources $c_s$ and $c_a$ are positive angular integrals that depend on the stiffness anisotropy, with $c_s=c_a=4\pi$ at isotropy (Appendix~\ref{app:rg}). The stiffnesses flow under two competing effects. Vortex dipoles screen them: elementary vortices act on both sectors, same-vorticity pairs only on the common sector, and opposite-vorticity pairs only on the relative sector. Josephson dipoles renormalize the relative sector through kernels modulated by the soliton lattice; these increase the relative stiffness at zero modulation and are cut off once $2\pi\rho_{\rm sol}a_0e^\ell$ exceeds order one. The flows generate anisotropy in both sectors, while an initially isotropic theory stays isotropic at zero density. The explicit Josephson flow holds at weak coupling below this modulation scale. Beyond it, a sliding state is matched to a theory without the oscillatory vertex, whereas a state in which locking reaches strong coupling first requires a locked-mode description. The matching parameters are nonuniversal, but the critical criteria follow from defect marginality.
	
	\emph{C side.} At $\rho_{\rm sol}=0$, $y_J$ flows to strong coupling and pins $\theta_-$. The massive relative mode confines elementary vortices and opposite-vorticity pairs, leaving $y_s$, which becomes marginal at $2-4\pi\mathcal K_{\rm c}=0$. With $\mathcal K_{\rm c}=\Upsilon_{+,g}/T$, this reproduces Eq.~\eqref{eq:lockedcriterion} and the universal exponent $\eta_l=1/4$.
	
		\emph{IC side.} In the matched sliding theory, both phase modes remain massless. When both pair-vortex channels are irrelevant, $y_1$ becomes marginal at $2-\pi(\mathcal K_{\rm c}+\mathcal K_g)=0$, which gives Eq.~\eqref{eq:singlecriterion} with the stiffness identifications above. Density-dependent matching and subsequent vortex screening select the limiting ratio $\mathcal K_g/\mathcal K_{\rm c}$ and hence the layer exponent. In the $(\mathcal K_{\rm c},\mathcal K_g)$ plane, the transition thus lies on a two-mode fixed line, a one-parameter family of BKT fixed points; the full Gaussian theory also retains spatial metric parameters.

		\emph{Dilute-PT stability.} For dilute noncrossing walls with short-range interactions and finite positive tilt stiffness, the energy cost of tilting a wall, thermal PT theory gives $\mathcal K_g\to1/(4\pi)$ at fixed $T>0$ (Appendix~\ref{app:rg}). Equation~\eqref{eq:compactRG} then gives the $y_a$ eigenvalue $2-4\pi\mathcal K_g\to1>0$: opposite-vorticity pairs are relevant, and the source $c_ay_1^2$ generates them even from zero initial fugacity. As for dislocations near 2D commensurate--incommensurate transitions~\cite{Coppersmith1981}, the elementary-vortex critical segment therefore cannot extend to the dilute PT limit. Because these pairs carry no common winding, the instability alone neither fixes the replacement phase nor implies melting into L.

	\subsection{Soliton entry at finite temperature}

		To test soliton entry at finite temperature, we compare neighboring fixed-soliton sectors, because local updates rarely change the number of complete soliton walls. Thermodynamic integration gives the free-energy difference per unit wall length between sectors with $N+1$ and $N$ solitons, $\Delta f_N(k_0)$, and the crossing $\Delta f_N(k_N)=0$ defines the field $k_N$ at which sectors $N$ and $N+1$ exchange stability. No explicit winding constraint is imposed, so sector retention relies on metastability; Appendix~\ref{app:mc} reports the available winding diagnostics. With open boundaries along $x$ and $L_x$ sites in that direction, we assign each crossing the midpoint density $\rho_{N+1/2}=(N+1/2)/(L_x-1)$. PT scaling then predicts $k_N-\kc(T)\propto\rho_{N+1/2}^2$, so the finite-temperature threshold $\kc(T)$ appears as the zero-density intercept. Forward and backward scans monitor hysteresis, and statistical errors are propagated through the integration.
	
	\begin{figure}[tb]
		\centering
		\includegraphics[width=\linewidth]{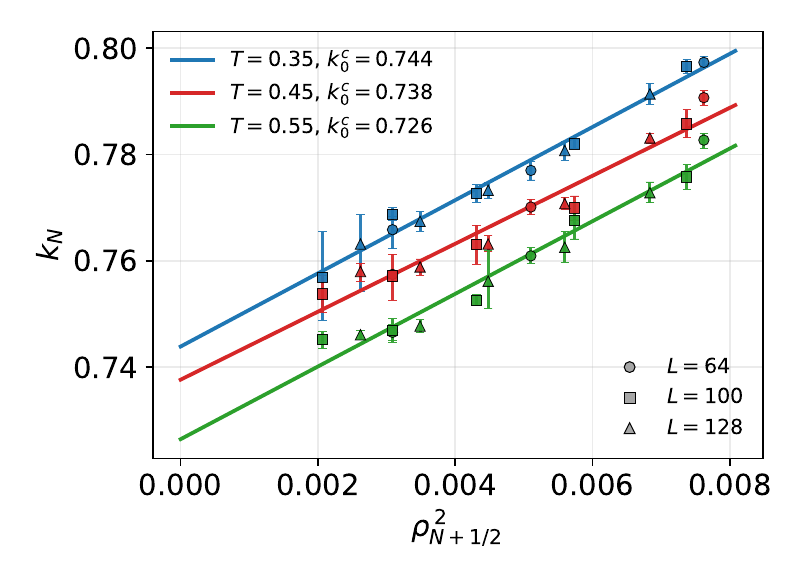}
			\caption{Finite-temperature PT scaling of fixed-soliton-sector crossings at $K=0.2$. The crossing field $k_N$ between sectors $N$ and $N+1$ is plotted against $\rho_{N+1/2}^2$, with $\rho_{N+1/2}=(N+1/2)/(L_x-1)$, for $T=0.35$, 0.45, and 0.55 (colors) and linear sizes $L=64$, 100, and 128 (symbols); error bars are conditional statistical 95\% intervals. Lines show fits to $k_N=\kc(T)+A(T)\rho_{N+1/2}^2$ over $0.002\le\rho_{N+1/2}^2\le0.008$, with intercepts $\kc(T)$. This linear dependence corresponds to the PT onset $\rho_{\rm sol}\propto[k_0-\kc(T)]^{1/2}$.}
		\label{fig:ci}
	\end{figure}
	
	Figure~\ref{fig:ci} shows that $k_N$ depends linearly on $\rho_{N+1/2}^2$ over the sampled densities, as PT scaling predicts. The fits give $\kc=0.7438$, $0.7376$, and $0.7264$ at $T=0.35$, $0.45$, and $0.55$, with statistical 95\% intervals $[0.7394,0.7487]$, $[0.7351,0.7398]$, and $[0.7252,0.7278]$, respectively, conditional on a unique crossing in each bootstrap resample (Appendix~\ref{app:mc}). Integration-rule, fit-range, and scan-direction sensitivities, reported separately in Table~\ref{tab:kc-sensitivity}, are as large as $0.011$ and exceed these statistical intervals, which therefore do not represent total uncertainties.
	
	\subsection{Thermal melting}
		We test the two melting criteria on thermal cuts through the C and IC phases, locating C--L melting from Eq.~\eqref{eq:lockedcriterion} and IC melting from Eq.~\eqref{eq:singlecriterion} with the helicity moduli defined above. As an independent probe, we compute the layer correlations
	$C_{ab}(r)=\left\langle e^{i[\theta_a(\mathbf{R}+r\hat y)-\theta_b(\mathbf{R})]}\right\rangle$, with $a,b=1,2$,
		averaged over $\mathbf{R}$. The diagonal correlations $C_{11}$ and $C_{22}$ give the physical-layer exponent $\eta_l$, expected to be $1/4$ at C--L melting and $\eta_l^*$, set by $\Upsilon_{-,g}/\Upsilon_{+,g}$, at IC melting; $C_{12}$ and $C_{21}$ provide complementary interlayer-coherence diagnostics. Finite-size tests and fitting details are given in Appendix~\ref{app:mc}.
	
		\begin{figure}[!t]
			\centering
			\begin{minipage}{0.96\linewidth}
				\centering
				\begin{overpic}[width=\linewidth]{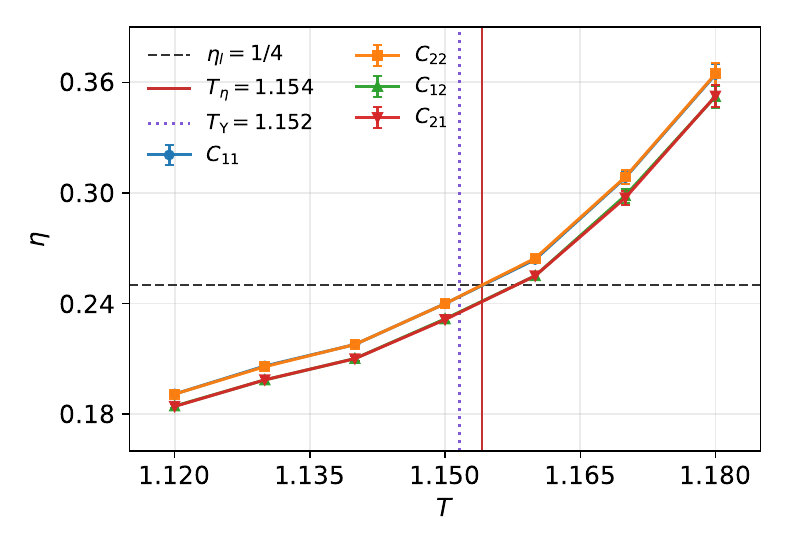}
					\put(2,65){\bfseries (a)}
				\end{overpic}
			\end{minipage}
			\hfill
			\begin{minipage}{0.96\linewidth}
				\centering
				\begin{overpic}[width=\linewidth]{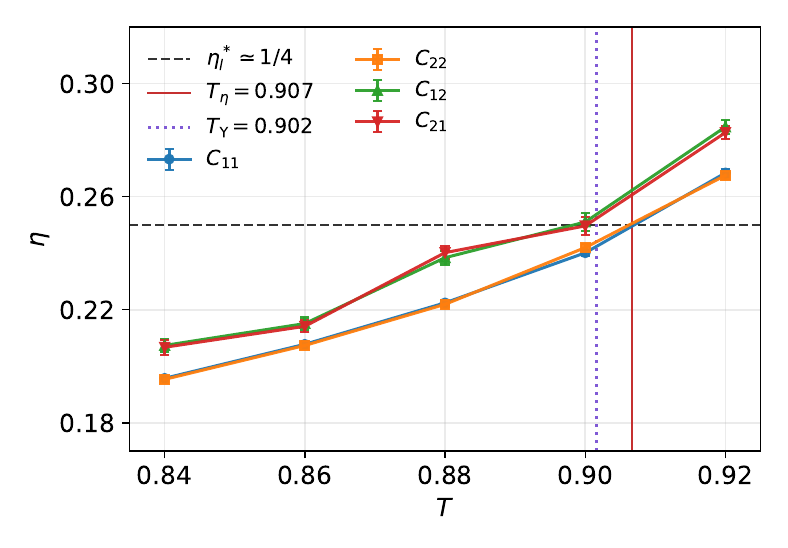}
					\put(2,65){\bfseries (b)}
				\end{overpic}
			\end{minipage}
				\caption{Correlation exponents at thermal melting for $K=0.2$ and $L=100$: (a) C--L cut at $k_0=0.25$; (b) IC--L cut at $k_0=1.5$. Symbols show exponents $\eta_{ab}$ from fits $|C_{ab}(r)|\propto r^{-\eta_{ab}}$, with bootstrap error bars. The dashed line marks $\eta_l=1/4$ in (a) and the stiffness-ratio prediction $\eta_l^*\simeq1/4$ in (b). Solid vertical lines mark $T_\eta$, where $(\eta_{11}+\eta_{22})/2=1/4$, and dotted lines mark $T_\Upsilon$ from Eq.~\eqref{eq:lockedcriterion} in (a) and Eq.~\eqref{eq:singlecriterion} in (b); in (b), $T_\eta$ is shown only for comparison. The off-diagonal $C_{12}$ and $C_{21}$ provide complementary interlayer-coherence diagnostics. C--L finite-size results are given in Appendix~\ref{app:mc}.}
			\label{fig:eta}
		\end{figure}
	
	Figure~\ref{fig:eta} compares the two diagnostics on representative $L=100$ cuts. On the C--L cut, the diagonal exponents reach $1/4$ at $T_\eta=1.154$, close to $T_\Upsilon=1.152$ from Eq.~\eqref{eq:lockedcriterion}. On the IC--L cut, they cross $1/4$ at $T_\eta=0.907$, near $T_\Upsilon=0.902$ from Eq.~\eqref{eq:singlecriterion}, where the nearly equal common and relative helicity moduli give $\eta_l^*\simeq1/4$. Each cut is thus consistent with its own melting criterion.
	At the interpolated IC crossing, $T_\Upsilon=0.9016(8)$, we obtain $\Upsilon_{+,g}=0.2872(5)$, $\Upsilon_{-,g}=0.2868(6)$, and $r=\Upsilon_{-,g}/\Upsilon_{+,g}=0.9988(35)$, with one-standard-deviation uncertainties from 600 paired whole-chain bootstrap resamples. Both moduli exceed $T_\Upsilon/(2\pi)=0.14350(12)$, and the 95\% interval $0.9915<r<1.0048$ lies inside $1/3<r<3$. With $\mathcal K_g=\Upsilon_{-,g}/T$, these estimates place both pair-vortex channels in the irrelevant range of the weak-fugacity RG criterion at this finite soliton density, as required for elementary-vortex melting.

	\section{Discussion and conclusion}
	Interlayer phase locking and sliding select which vortices melt a superconducting bilayer. Locking confines every defect with relative winding, so same-vorticity layer pairs melt the C phase; sliding frees elementary layer vortices, which melt a stable IC phase. Fixed-sector crossings show PT scaling of soliton entry, and at finite soliton density the measured stiffnesses place both pair-vortex channels in their irrelevant range. Near entry, however, opposite-vorticity pairs destabilize the dilute sliding state, leaving the equilibrium phase structure there open.

	The two melting channels have distinct experimental signatures. Resistance measurements can track the C--L and IC--L melting lines through the Halperin--Nelson form $R(T)\!\propto\!\exp[-b/\sqrt{T-T_{\rm BKT}}]$, with nonuniversal $b$ and melting temperature $T_{\rm BKT}$~\cite{HalperinNelson1979,Beasley1979,Minnhagen1987}. Penetration-depth or mutual-inductance measurements, which probe the common-mode superfluid stiffness, can distinguish the two melting channels through its universal jump: Eqs.~\eqref{eq:lockedcriterion} and \eqref{eq:singlecriterion} give $\Upsilon_{+,g}/T=1/(2\pi)$ at C--L melting but $2/[\pi(1+r)]$ at IC--L melting, larger by $4/(1+r)\simeq2$ for $r\simeq1$. Phase-sensitive probes distinguish the ordered phases: the FF state has predominantly zero-momentum interlayer coherence, whereas the Bloch SC soliton lattice produces finite-momentum components of $e^{i\theta_-}$ at wave vector $2\pi\rho_{\rm sol}$ and its harmonics. Momentum-resolved interlayer tunneling or patterned Josephson interferometry could therefore detect soliton entry and track the soliton density as a function of field.

	Qiu and Zhou identified the zero-temperature FF--Bloch SC transition as a PT transition~\cite{QiuZhou2022}, and Nag \emph{et al.}\ supported its ordered-state structure with BdG calculations~\cite{Nag2025}. Our results extend this zero-temperature picture to thermal melting. The locking--sliding mechanism offers a framework for other coupled compact $U(1)$ systems in which commensurate--incommensurate ordering changes defect confinement. Disorder, dynamical tunneling signatures, and multilayer systems are natural directions for further study.

	\section*{Conflict of interest}
	The authors declare that they have no conflict of interest.

	\section*{Acknowledgments}
	This work was supported by the National Key Research and Development Program of China (Grant No. 2022YFA1403403) and the National Natural Science Foundation of China (Grant Nos. 12274441 and 12534004).

	\section*{Author contributions}
	Yi Zhou conceived the project and developed the theory. Yicheng Zhong performed the numerical calculations and contributed to the renormalization-group analysis. Both authors analyzed the results and wrote the manuscript.

	\section*{Data availability}
	The data supporting the findings of this study are available from the corresponding author upon reasonable request.

	\section*{Declaration of generative AI and AI-assisted technologies in the manuscript preparation process}
	During the preparation of this work, the authors used Claude Opus 5.5 (Anthropic), through the Claude Code interface, in order to revise and restructure the manuscript text, check the consistency of equations, notation, and references, draft wording for the authors' review, refine the TikZ code of Fig.~\ref{fig:model}, and prepare the LaTeX files. The authors also used OpenAI's GPT-6-Astra and GPT-5.6-Sol, through the Codex interface, in order to develop simulation and analysis code, analyze the numerical data, and perform Wolfram Language checks of the equations. These tools were also used to check and develop parts of the analytical derivations. After using these tools, the authors reviewed and edited the content as needed and take full responsibility for the content of the published article.

	\onecolumngrid
\appendix
\setcounter{secnumdepth}{2}
\section{Renormalization-group analysis}
\label{app:rg}

This section derives the defect marginality conditions used in the main text within a weak-fugacity renormalization-group (RG) scheme. The microscopic variables are the compact layer phases $\theta_1$ and $\theta_2$; we use $\phic=\theta_1+\theta_2$ and $\phir=\theta_1-\theta_2$ as long-wavelength fields while retaining the vortex content inherited from the layer phases. The long-distance phase stiffnesses in these conditions are distinct from the finite-size helicity moduli used in the numerical comparison below. As in the main text, C/FF and IC/Bloch SC denote the locked commensurate and sliding incommensurate phases, and L the thermal liquid.

We set $k_{\rm B}=1$. In the fixed-$k_0$ (grand-canonical) formulation, the isotropic continuum action is
\begin{equation}
S=\int d^2r\,\left[\frac{\mathcal K_{\rm c}}{2}(\nabla\phic)^2+\frac{\mathcal K_{\rm r}}{2}(\nabla\phir-k_0\hat{x})^2-\frac{y_J}{a_0^2}\cos\phir\right],
\end{equation}
where $\mathcal K_{\rm c}$ and $\mathcal K_{\rm r}$ are the dimensionless common and relative stiffnesses and $k_0$ is the field-induced momentum mismatch. With this convention, the Josephson fugacity is $y_J(0)=Ka_0^2/T$ at the microscopic cutoff $a_0$, where the RG scale $\ell$ starts and $K$ is the phase-only Josephson coupling per unit area. Expanding the relative-gradient term and dropping the constant $\mathcal K_{\rm r}k_0^2L_xL_y/2$ for a system of area $L_xL_y$ gives
\begin{equation}
S_{\rm gc}=\frac{1}{2}\int d^2r\,\left[\mathcal K_{\rm c}(\nabla\phic)^2+\mathcal K_{\rm r}(\nabla\phir)^2\right]-\frac{y_J}{a_0^2}\int d^2r\,\cos\phir-\mathcal K_{\rm r} k_0\int d^2r\,\partial_x\phir ,
\end{equation}
so $k_0$ acts as a chemical potential conjugate to the soliton density
\begin{equation}
\rho_{\rm sol}=\frac{1}{2\pi L_xL_y}\int d^2r\,\partial_x\phir .
\end{equation}

To separate the uniform soliton background from its fluctuations, we follow the standard fixed-density RG treatment of the finite-density Pokrovsky--Talapov (PT) model~\cite{PokrovskyTalapov1979,Coppersmith1981,Horovitz1983} and write
\begin{equation}
\phir(x,y)=\varphi(x,y)+Qx,\qquad
Q=2\pi\rhosol,
\end{equation}
where $\varphi$ is periodic and $\rhosol$ is held fixed during the RG. The action becomes
\begin{equation}
S_{\rm c}=\frac{1}{2}\int d^2r\,\left[\mathcal K_{\rm c}(\nabla\phic)^2+\mathcal K_{\rm r}(\nabla\varphi)^2\right]-\frac{y_J}{a_0^2}\int d^2r\,\cos(\varphi+Qx)+\frac{\mathcal K_{\rm r}}{2}(Q-k_0)^2L_xL_y ,
\end{equation}
since the cross term proportional to $(Q-k_0)\int d^2r\,\partial_x \varphi$ vanishes for periodic $\varphi$. At the starting scale, $k_0$ thus enters only through the background free energy $\mathcal K_{\rm r}(Q-k_0)^2L_xL_y/2$, which selects the physical soliton density by comparing fixed-density sectors; the fluctuating fields are then coarse grained at fixed $\rhosol$.

Finite soliton density makes the relative Gaussian sector anisotropic, and vortex screening can transmit this anisotropy to the common sector. We therefore retain directional stiffnesses in both sectors:
\begin{equation}
S_0=\frac{1}{2}\int d^2r\,\sum_{\nu=x,y}\left[\mathcal K_{{\rm c},\nu}(\partial_\nu\phic)^2+\mathcal K_{{\rm r},\nu}(\partial_\nu\varphi)^2\right],
\label{eq:sm-Gaussian}
\end{equation}
with geometric means $\mathcal K_{\rm c}=\sqrt{\mathcal K_{{\rm c},x}\mathcal K_{{\rm c},y}}$ and $\mathcal K_g=\sqrt{\mathcal K_{{\rm r},x}\mathcal K_{{\rm r},y}}$. In the relative sector we parameterize $\mathcal K_{{\rm r},y}=\mathcal K_{\rm r}$ and $\mathcal K_{{\rm r},x}=\mathcal K_{\rm r}(1-\zeta)$, so $\mathcal K_g=\mathcal K_{\rm r}\sqrt{1-\zeta}$. All four stiffnesses are positive, so $\mathcal K_{\rm r}>0$ and $\zeta<1$; their microscopic values are isotropic, $\rho_s/(2T)$, with $\rho_s$ the bare single-layer stiffness, and isotropy persists in the C limit $\rhosol=0$.

The shift $\phir=\varphi+Qx$ treats the smooth soliton background but leaves the vortex content unchanged: the compact variables remain the layer phases $\theta_1$ and $\theta_2$, and vortices are labeled by layer windings $(w_1,w_2)\in\mathbb{Z}^2$, or equivalently by the induced common and relative windings. Because $\phic$ and $\phir$ are not independently compact, an elementary layer vortex necessarily carries both. In terms of the dual fields $\vartheta_1$ and $\vartheta_2$, with $\vartheta_c=\vartheta_1+\vartheta_2$ and $\vartheta_r=\vartheta_1-\vartheta_2$, the elementary-vortex action is
\begin{equation}
S_1=-\frac{2y_1}{a_\ell^2}\int d^2r\,\left[\cos(2\vartheta_1)+\cos(2\vartheta_2)\right]=-\frac{2y_1}{a_\ell^2}\int d^2r\,\left[\cos(\vartheta_c+\vartheta_r)+\cos(\vartheta_c-\vartheta_r)\right].
\end{equation}
Correlated same- and opposite-vorticity pairs, defined in the main text, contribute
\begin{equation}
S_s=-\frac{y_s}{a_\ell^2}\int d^2r\,\cos(2\vartheta_c),\qquad S_a=-\frac{y_a}{a_\ell^2}\int d^2r\,\cos(2\vartheta_r),
\end{equation}
where $a_\ell=a_0e^\ell$ is the running cutoff. The negative cosine signs correspond to positive vortex activities; with these prefactors, a signed elementary charge has activity $y_1$, and signed same- and opposite-vorticity pairs have $y_s/2$ and $y_a/2$. The Josephson perturbation keeps the form $-y_Ja_\ell^{-2}\int d^2r\,\cos(\varphi+Qx)$. The phase and dual operators perturb the same Gaussian theory; they are not independent copies of its degrees of freedom.

We coarse grain with circular real-space separation shells $a_\ell<|r|<a_\ell e^{d\ell}$, fixed laboratory axes, and isotropic coordinate rescaling. This positive-activity convention is related to, but not identical with, the amplitude conventions of Refs.~\cite{Benfatto2007,Mathey2008}. Close nonneutral clusters are absorbed into composite-charge activities and are no longer counted as resolved pairs below the new cutoff, while microscopic core contributions enter the initial activities. Gaussian vertices are normal ordered with the determinant-one distance metrics $R_\alpha^2=|r|^2s_\alpha(\theta)$, where $\theta$ is the polar angle of $\mathbf r$ and
\begin{equation}
s_\alpha(\theta)=
\sqrt{\frac{\mathcal K_{\alpha,y}}{\mathcal K_{\alpha,x}}}\cos^2\theta+
\sqrt{\frac{\mathcal K_{\alpha,x}}{\mathcal K_{\alpha,y}}}\sin^2\theta,
\qquad \alpha={\rm c,r}.
\label{eq:sm-RG-metrics}
\end{equation}
Fusion of charges $q=(q_c,q_r)$ and $q'=(q'_c,q'_r)$ carries the angular weight $s_{\rm c}^{\pi\mathcal K_{\rm c}q_cq'_c}s_{\rm r}^{\pi\mathcal K_g q_rq'_r}$. Summing ordered charge pairs with a factor $1/2$ gives the fugacity flows to second order:
\begin{align}
\frac{dy_J}{d\ell}&=\left(2-\frac{1}{4\pi \mathcal K_g}\right)y_J, \label{eq:sm-RG-yJ}\\
\frac{dy_1}{d\ell}&=\left[2-\pi\left(\mathcal K_{\rm c}+\mathcal K_g\right)+b_s y_s+b_a y_a\right]y_1, \\
\frac{dy_s}{d\ell}&=\left(2-4\pi \mathcal K_{\rm c}\right)y_s+c_s y_1^2, \label{eq:sm-RG-ys}\\
\frac{dy_a}{d\ell}&=\left(2-4\pi \mathcal K_g\right)y_a+c_a y_1^2 .
\label{eq:sm-fugacity-RG}
\end{align}
The coefficients are positive angular integrals of the running Gaussian metrics,
\begin{align}
b_s&=\frac12\int_0^{2\pi}d\theta\,s_{\rm c}^{-2\pi\mathcal K_{\rm c}},
& b_a&=\frac12\int_0^{2\pi}d\theta\,s_{\rm r}^{-2\pi\mathcal K_g}, \label{eq:sm-RG-fusion-b}\\
c_s&=2\int_0^{2\pi}d\theta\,s_{\rm c}^{\pi\mathcal K_{\rm c}}s_{\rm r}^{-\pi\mathcal K_g},
& c_a&=2\int_0^{2\pi}d\theta\,s_{\rm c}^{-\pi\mathcal K_{\rm c}}s_{\rm r}^{\pi\mathcal K_g},
\label{eq:sm-RG-fusion-c}
\end{align}
which reduce to $b_s=b_a=\pi$ and $c_s=c_a=4\pi$ at isotropy, where $s_{\rm c}=s_{\rm r}=1$. The pair-vortex sources count composite charges in the coarse-grained gas and need not equal connected interlayer cosine couplings in other RG conventions. All flows omit terms of third and higher total order in the small couplings.

Vortex screening is governed by the dual Gaussian action
\begin{equation}
S_{0,d}=\frac{1}{8\pi^2}\int d^2r\,\sum_{\alpha={\rm c,r}}
\left[\frac{(\partial_x\vartheta_\alpha)^2}{\mathcal K_{\alpha,y}}
+\frac{(\partial_y\vartheta_\alpha)^2}{\mathcal K_{\alpha,x}}\right],
\label{eq:sm-RG-dual}
\end{equation}
in which duality exchanges the directional inverse stiffnesses and thereby fixes which angular moment screens each phase stiffness. A charge $q$ has scaling dimension $\pi(q_c^2\mathcal K_{\rm c}+q_r^2\mathcal K_g)$, and the neutral-dipole angular weights and their moments are
\begin{equation}
W_1=s_{\rm c}^{-\pi\mathcal K_{\rm c}}s_{\rm r}^{-\pi\mathcal K_g},\qquad
W_s=s_{\rm c}^{-4\pi\mathcal K_{\rm c}},\qquad
W_a=s_{\rm r}^{-4\pi\mathcal K_g},
\label{eq:sm-RG-vortex-weights}
\end{equation}
\begin{equation}
A_{\mu x}=\frac1\pi\int_0^{2\pi}d\theta\,\sin^2\theta\,W_\mu,\qquad
A_{\mu y}=\frac1\pi\int_0^{2\pi}d\theta\,\cos^2\theta\,W_\mu,
\qquad \mu=1,s,a.
\label{eq:sm-RG-vortex-moments}
\end{equation}
All $A_{\mu\nu}$ are positive and equal one at isotropy. Equal elementary-layer activities cancel mixed common--relative screening, and reflection symmetry cancels spatial off-diagonal terms.

For Josephson dipoles, we define $p_\ell=Qa_\ell=2\pi\rhosol a_\ell$, which obeys $dp_\ell/d\ell=p_\ell$, and $e_x(\theta)=\cos\theta$, $e_y(\theta)=\sin\theta$. The neutral electric pair, a Josephson dipole, has weight $s_{\rm r}^{-1/(4\pi\mathcal K_g)}$, retains the modulation factor $\cos(p_\ell\cos\theta)$, and enters with the cosine-product coefficient $y_J^2/4$. Expanding to second order in the slow phase gradient gives the kernels
\begin{equation}
B_\nu(p)=\frac1\pi\int_0^{2\pi}d\theta\,e_\nu^2(\theta)
s_{\rm r}^{-1/(4\pi\mathcal K_g)}\cos(p\cos\theta),\qquad \nu=x,y,
\label{eq:sm-RG-Josephson-kernels}
\end{equation}
which, unlike the vortex moments, use the unexchanged spatial directions. For an isotropic relative sector,
\begin{equation}
B_x(p)=J_0(p)-J_2(p),\qquad B_y(p)=J_0(p)+J_2(p),
\label{eq:sm-RG-Josephson-isotropic}
\end{equation}
where $J_n$ are Bessel functions, so $B_x(0)=B_y(0)=1$: at zero modulation, Josephson dipoles increase both relative stiffnesses, whereas the finite-modulation kernels need not remain positive. The same expansion also produces a background free-energy term and a term linear in $\partial_x\varphi$, which integrates to zero for the smooth periodic fluctuation field at fixed density. Sector free energies require their own treatment and are not determined by the stiffness flows alone.

Combining vortex and Josephson dipoles in this separation-shell convention gives the stiffness flows
\begin{align}
\frac{d\mathcal K_{{\rm c},\nu}}{d\ell}
&=-4\pi^3\mathcal K_{{\rm c},\nu}^2
\left(2A_{1\nu}y_1^2+A_{s\nu}y_s^2\right), \label{eq:sm-RG-Kc}\\
\frac{d\mathcal K_{{\rm r},\nu}}{d\ell}
&=\frac{\pi y_J^2}{4}B_\nu(p_\ell)
-4\pi^3\mathcal K_{{\rm r},\nu}^2
\left(2A_{1\nu}y_1^2+A_{a\nu}y_a^2\right),\qquad \nu=x,y,
\label{eq:sm-RG-Kr}
\end{align}
where the factor two counts the two elementary-layer species and, for a pair, the squared charge of four cancels the factor $1/4$ from its squared signed activity. Same-vorticity pairs screen only the common sector, opposite-vorticity pairs only the relative sector, and elementary vortices both. These powers and coefficients belong to the stated positive-activity convention; they are not obtained by changing individual powers in the amplitude equations of Ref.~\cite{Mathey2008}.

The relative anisotropy follows from the directional flows rather than from an independent Josephson-only equation:
\begin{equation}
\frac{d\zeta}{d\ell}
=-\frac{1}{\mathcal K_{{\rm r},y}}\frac{d\mathcal K_{{\rm r},x}}{d\ell}
+\frac{\mathcal K_{{\rm r},x}}{\mathcal K_{{\rm r},y}^2}
\frac{d\mathcal K_{{\rm r},y}}{d\ell},
\label{eq:sm-stiffness-RG}
\end{equation}
which includes vortex-induced relative anisotropy. Elementary-vortex dipoles can also generate common anisotropy when the two sector metrics differ; Eqs.~\eqref{eq:sm-RG-Kc}--\eqref{eq:sm-RG-Kr} retain both effects.

We use the explicit modulated Josephson flow at weak coupling while $p_\ell\lesssim1$ and do not extrapolate the sharp-shell Bessel kernels to large $p_\ell$ with a growing local $y_J$. If the flow reaches a sliding IC state with weak Josephson coupling, we match to a long-distance theory without the oscillatory vertex and continue the vortex flows from the matched stiffnesses and activities; the matching contributions are nonuniversal. If locking instead reaches strong coupling first, a locked-relative-mode description replaces the perturbative flow. Finite density alone therefore does not guarantee a stable sliding state.

The retained equations are a second-order projection, not a closed algebra of all operators. Quadratic fusion also generates higher vortex charges, which are irrelevant on the open elementary-vortex critical segment where both pair-vortex channels are irrelevant. Electric (Josephson) harmonics and mixed electric--magnetic (Josephson--vortex) operators are generated as well; if initially absent as independent leading perturbations, they feed back into the retained equations only at third or higher order. The critical analysis below uses the weak-coupling matched theory rather than a continuation through strong coupling.

\subsection{C--L critical point and layer-correlation exponent}

We first extract the C--L critical behavior from the same RG equations. In the commensurate phase, $\rhosol=0$ and $\zeta=0$, so $\mathcal K_g=\mathcal K_{\rm r}$ and the Josephson flow is
\begin{equation}
\frac{dy_J}{d\ell}
=\left(2-\frac{1}{4\pi\mathcal K_{\rm r}}\right)y_J.
\end{equation}
In the ordered C phase, $y_J$ flows to strong coupling and pins $\phir$, so the relative mode is massive beyond the locking length. Elementary layer vortices and opposite-vorticity pairs, which carry relative winding, are then confined and do not survive as asymptotic defects.

The remaining BKT defect is the same-vorticity layer pair, with fugacity $y_s$. With the confined elementary-vortex fugacity set to zero, Eq.~\eqref{eq:sm-RG-ys} gives
\begin{equation}
\frac{dy_s}{d\ell}
=\left(2-4\pi\mathcal K_{\rm c}\right)y_s+o(y_s),
\end{equation}
so marginality of this defect requires
\begin{equation}
\mathcal K_{\rm c}^*=\frac{1}{2\pi}.
\label{eq:sm-CL-marginality}
\end{equation}
Consistently, once the relative sector is locked, the isotropic common-stiffness flow in Eq.~\eqref{eq:sm-RG-Kc} reduces to
\begin{equation}
\frac{d\mathcal K_{\rm c}}{d\ell}
=-4\pi^3\mathcal K_{\rm c}^2y_s^2.
\end{equation}
Along the critical separatrix, $y_s\rightarrow0$ and $\mathcal K_{\rm c}\rightarrow1/(2\pi)$, giving the usual isolated BKT critical point. With $\mathcal K_{\rm c}=\Upsilon_{+,g}/T$, where $\Upsilon_{+,g}$ is the geometric-mean common helicity modulus defined in the helicity-modulus subsection below, this is equivalent to
\begin{equation}
\Upsilon_{+,g}^*=\frac{T_c}{2\pi}
\end{equation}
at the critical temperature $T_c$.

Josephson locking also sets $\theta_1=\theta_2=\phic/2$ up to massive relative-phase fluctuations, so the diagonal physical-layer correlation $C_{ll}(r)$ of Eq.~\eqref{eq:sm-Cab} obeys
\begin{equation}
C_{ll}(r)\sim r^{-\eta_l^*},\qquad
\eta_l^*=\frac{1}{8\pi\mathcal K_{\rm c}^*}=\frac{1}{4},
\label{eq:sm-CL-eta}
\end{equation}
a unique universal exponent at the C--L transition. By contrast, at an elementary-vortex-driven IC--L transition with irrelevant pair-vortex channels, both the common and the sliding relative modes remain massless, and marginality fixes only the sum of their stiffnesses.

\subsection{IC--L critical line and layer-correlation exponent}

We extract the IC--L critical behavior from Eqs.~\eqref{eq:sm-RG-yJ}--\eqref{eq:sm-fugacity-RG} and \eqref{eq:sm-RG-Kc}--\eqref{eq:sm-RG-Kr} in the matched sliding theory. For $\rhosol>0$, $p_\ell=2\pi\rhosol a_0e^\ell$ grows under coarse graining; once the sliding state is matched across the modulation scale, the long-distance theory retains two massless Gaussian modes and the vortex perturbations but no oscillatory Josephson vertex.

Consider an IC--L transition driven by elementary layer vortices. When the pair-vortex fugacities $y_s$ and $y_a$ are irrelevant, their $O(y_1^2)$ sources vanish asymptotically together with $y_1$, and near the critical fixed point the $y_1$ flow reduces to
\begin{equation}
\frac{dy_1}{d\ell}
=\left[2-\pi\left(\mathcal K_{\rm c}+\mathcal K_g\right)\right]y_1+o(y_1),
\end{equation}
so marginality of the elementary layer vortex requires
\begin{equation}
\mathcal K_{\rm c}^*+\mathcal K_g^*=\frac{2}{\pi}.
\label{eq:sm-IC-marginality}
\end{equation}
On the critical trajectory of this matched theory, the vortex fugacities flow to zero, and the directional stiffness beta functions in Eqs.~\eqref{eq:sm-RG-Kc}--\eqref{eq:sm-RG-Kr} vanish asymptotically. Because Eq.~\eqref{eq:sm-IC-marginality} fixes only one combination of the two limiting geometric-mean stiffnesses, the RG transition terminates, in the $(\mathcal K_{\rm c},\mathcal K_g)$ plane, on a two-mode fixed line, a one-parameter family of BKT fixed points; the full Gaussian fixed-point theory also retains spatial metric parameters.

At a point on this fixed line, the physical-layer operator is
\begin{equation}
e^{i\theta_l}=e^{\pm iQx/2}\exp\!\left[\frac{i}{2}\left(\phic\pm\varphi\right)\right],
\end{equation}
where the upper (lower) sign refers to layer 1 (2). The nonfluctuating background factor cancels in diagonal correlations along $y$, and the scaling dimension of the remaining vertex operator follows from the fixed-point Gaussian action:
\begin{equation}
C_{ll}(r_y)\sim r_y^{-\eta_l^*},\qquad
\eta_l^*=\frac{1}{8\pi}\left(\frac{1}{\mathcal K_{\rm c}^*}+\frac{1}{\mathcal K_g^*}\right).
\label{eq:sm-IC-eta}
\end{equation}
If the relative helicity modulus equals the relative-phase stiffness, $\mathcal K_g=\Upsilon_{-,g}/T$, then with $\mathcal K_{\rm c}=\Upsilon_{+,g}/T$ this becomes
\begin{equation}
\eta_l^*=\frac{T}{8\pi}\left(\frac{1}{\Upsilon_{+,g}^*}+\frac{1}{\Upsilon_{-,g}^*}\right).
\end{equation}

With $r=\mathcal K_g^*/\mathcal K_{\rm c}^*$, Eq.~\eqref{eq:sm-IC-marginality} gives
\begin{equation}
\eta_l^*=\frac{(1+r)^2}{16r}
=\frac{1}{4}+\frac{(r-1)^2}{16r},
\label{eq:sm-IC-eta-ratio}
\end{equation}
so $\eta_l^*=1/4$ at the symmetric point $r=1$ and stays close to $1/4$ when the two stiffnesses are comparable.

Irrelevance of both pair-vortex channels requires negative linear eigenvalues in the $y_s$ and $y_a$ flows,
\begin{equation}
2-4\pi\mathcal K_{\rm c}^*<0,\qquad
2-4\pi\mathcal K_g^*<0,
\end{equation}
which, together with Eq.~\eqref{eq:sm-IC-marginality}, give
\begin{equation}
\frac{1}{3}<r<3,\qquad
\frac{1}{4}\leq\eta_l^*<\frac{1}{3}.
\end{equation}
At either endpoint of this interval, one pair-vortex channel becomes marginal as well.

The soliton density enters through $p_\ell=2\pi\rhosol a_\ell$, which sets the modulation matching scale. The matched stiffnesses and activities, followed by vortex screening, help select the limiting ratio $r$; because these matching parameters are nonuniversal, the analysis does not imply a universal function $\eta_l^*(\rhosol)$. With $\mathcal K_{\rm c}=\Upsilon_{+,g}/T$, converting Eq.~\eqref{eq:sm-IC-marginality} into $\Upsilon_{+,g}+\Upsilon_{-,g}=2T/\pi$ additionally requires $\mathcal K_g=\Upsilon_{-,g}/T$, an identification that neither the Gaussian marginality calculation nor the present simulations test independently. We therefore use the IC helicity-modulus crossing as a finite-size melting diagnostic and treat the exponent prediction from the measured ratio, together with the pair-vortex inequalities, as comparisons with the sliding theory under this identification.

\subsection{Dilute-PT stability}

Pair-vortex irrelevance must also be checked near soliton entry. Consider the dilute-wall PT theory at fixed $T>0$, with noncrossing $2\pi$ soliton walls, short-range residual repulsion, no bound-wall formation, and no wall endpoints. Let $\tau>0$ be the finite wall tilt stiffness, defined by the energy cost $\tfrac{\tau}{2}\int dy\,(\partial_yu)^2$ for a displacement $u$ of one wall, and let $\mu$ be the excess wall chemical potential per unit length. The leading free-energy density is
\begin{equation}
f(\rhosol)=-\mu\rhosol+\frac{\pi^2T^2}{6\tau}\rhosol^3+o(\rhosol^3),
\label{eq:sm-PT-wall-free-energy}
\end{equation}
where the cubic term comes from thermal wandering of the noncrossing walls: their dilute transfer spectrum has kinetic eigenvalue $Tk^2/(2\tau)$, occupied up to $|k|=\pi\rhosol$, giving $T\int_{-\pi\rhosol}^{\pi\rhosol}(dk/2\pi)\,Tk^2/(2\tau)=\pi^2T^2\rhosol^3/(6\tau)$. Minimizing Eq.~\eqref{eq:sm-PT-wall-free-energy} to leading order gives $\mu=\pi^2T^2\rhosol^2/(2\tau)$, hence the PT square-root law when $\mu$ is linear in the field displacement from entry.

For a slowly varying wall displacement $u$, $\delta\rhosol=-\rhosol\partial_xu$ and the coarse-grained relative-phase fluctuation is $\varphi=-2\pi\rhosol u$, so compression and tilt cost
\begin{equation}
\delta F=\frac12\int d^2r\,\left[\rhosol^2 f''(\rhosol)(\partial_xu)^2+\tau\rhosol(\partial_yu)^2\right],
\label{eq:sm-PT-wall-elasticity}
\end{equation}
where $f''$ is taken with respect to $\rhosol$. Comparing $\delta F/T$ with Eq.~\eqref{eq:sm-Gaussian} gives the leading relative-phase stiffnesses
\begin{equation}
\mathcal K_{{\rm r},x}\simeq\frac{T\rhosol}{4\tau},\qquad
\mathcal K_{{\rm r},y}\simeq\frac{\tau}{4\pi^2T\rhosol},\qquad
\mathcal K_g\longrightarrow\frac{1}{4\pi}\quad(\rhosol\to0^+).
\label{eq:sm-PT-stiffness}
\end{equation}
These coefficients describe long-wavelength relative-phase fluctuations and are not obtained from the uniform relative helicity modulus. The limit $\rhosol\to0^+$ is taken after the thermodynamic limit at fixed $T>0$ and is distinct from the zero-temperature inverse-logarithmic onset.

The allowed opposite-vorticity layer pair, with $(w_+,w_-)=(0,2)$, has scaling dimension $4\pi\mathcal K_g\to1$, so Eq.~\eqref{eq:sm-fugacity-RG} gives the positive linear eigenvalue $2-4\pi\mathcal K_g\to1$; moreover, $c_ay_1^2$ generates this fugacity whenever elementary vortices have nonzero activity. The wall theory without endpoints is therefore unstable to these defects in the unrestricted compact model, independently of the common-phase stiffness. This is the bilayer counterpart of the dislocation instability of two-dimensional incommensurate phases near commensurate--incommensurate transitions~\cite{Coppersmith1981}. A fixed global soliton count alone does not exclude a neutral collection of local $(0,2)$ and $(0,-2)$ defects.

The fixed-sector crossings test PT scaling on the sampled scales, where a small defect activity need not yet have grown appreciably; they do not establish an asymptotically stable sliding phase arbitrarily close to entry. The pair-vortex check at $k_0=1.5$ instead probes a finite-density regime, where both pair-vortex channels are irrelevant under the identification $\mathcal K_g=\Upsilon_{-,g}/T$. Because opposite-vorticity pairs carry no common winding, this instability alone determines neither the fate of common-phase order nor a replacement phase boundary.

\section{Monte Carlo formulation and numerical analysis}
\label{app:mc}

\subsection{Lattice model and sampling}

On an $L\times L$ square lattice with unit lattice spacing we simulate
\begin{equation}
\mathcal H = -J\sum_{i,l}\cos(\theta_{l,i}-\theta_{l,i+\hat x}+s_lk_0/2)-J\sum_{i,l}\cos(\theta_{l,i}-\theta_{l,i+\hat y})-K\sum_i\cos(\theta_{1,i}-\theta_{2,i}).
\label{eq:sm-lattice-H}
\end{equation}
Here $i$ labels lattice sites, $l=1,2$ labels the layers, and $s_1=+1$, $s_2=-1$ make the field-induced momentum shifts opposite in the two layers. We set $J=1$, so that the bare single-layer stiffness is $\rho_s=J=1$ in lattice units, and $K$ is the Josephson coupling energy. The simulations update only the two compact layer phases $\theta_1$ and $\theta_2$; the common and relative fields are measured combinations, not independently simulated XY variables. One sweep consists of $2L^2$ attempted single-site Metropolis moves. The $L=64$, 100, and 128 thermal cohorts use $10^5$ equilibration and $3\times10^5$ measurement sweeps, averaged over 16 independent seeds for each parameter set.

For the C--L and IC--L boundary calculations, we use periodic boundaries in both directions to improve correlation-function sampling. The C cut is initialized in the locked sector. For the IC cut at $K=0.2$, $k_0=1.5$, and $L=100$, the sector is labeled by the integer windings $(W_1,W_2)$ of the layer phases around the periodic $x$ direction, with $N_{\rm sol}=W_1-W_2$ relative-phase solitons. A $y$-uniform zero-temperature energy scan selects $(W_1,W_2)=(12,-12)$, i.e.\ $N_{\rm sol}=24$, as the reference sector. To test this choice, we start four chains from the reference and from each of four neighboring sectors at $T=0.88$, $0.90$, and $0.92$, which bracket the IC--L crossing, using $10^5$ relaxation and $3\times10^5$ measurement sweeps. We record the integer layer windings during measurement and take the sector with the largest cumulative occupancy as dominant. It is $(12,-12)$ at all three temperatures; at $T=0.90$, some chains leave and re-enter this sector, and its occupancy is summed over repeated visits. All visited states are retained in the observables, and this operational selection does not determine equilibrium sector weights.

\subsection{Helicity moduli and correlations}

The helicity moduli are the free-energy curvatures per unit area with respect to uniform phase offsets $A$ in the intralayer cosine arguments, evaluated at $A=0$~\mbox{\cite{NelsonKosterlitz1977,WeberMinnhagen1988}}. For the common helicity modulus, we subtract $A/2$ from the arguments on all $\nu$ bonds ($\nu=x,y$) in both layers; for the relative one, we subtract $s_lA/2$ in layer $l$. The Josephson term is unchanged in both cases. With $C_{l\nu}=\sum_i\cos(\theta_{l,i}-\theta_{l,i+\hat\nu}+s_lk_0\delta_{\nu x}/2)$ and $S_{l\nu}=\sum_i\sin(\theta_{l,i}-\theta_{l,i+\hat\nu}+s_lk_0\delta_{\nu x}/2)$, the helicity moduli are
\begin{equation}
\Upsilon_{{\rm c/r},\nu}=\frac{J}{4L^2}\langle C_{1\nu}+C_{2\nu}\rangle-\frac{J^2}{4TL^2}\left[\langle(S_{1\nu}\!\pm\!S_{2\nu})^2\rangle-\langle S_{1\nu}\!\pm\!S_{2\nu}\rangle^2\right],
\label{eq:sm-helicity}
\end{equation}
where the upper (lower) sign gives the common (relative) helicity modulus, and the factor $1/4$, which fixes the normalization, follows from the increments of magnitude $A/2$ in the two layers.

The relative helicity modulus measures the curvature for these uniform phase offsets, whereas the relative-phase stiffness describes slow compression and tilt of the soliton lattice; the two coefficients need not coincide. For example, the relative $x$ offset replaces $k_0$ by $k_0-A$. In the smooth continuum at fixed integrated relative gradient $\int d^2r\,\partial_\nu{\phir}$, the free energy changes by a term linear in $A$ plus $\rho_s L_x L_y A^2/4$, giving a relative helicity modulus $\rho_s/2$, while the relative-phase stiffnesses depend on the soliton density. This continuum example does not replace the lattice estimator, but it shows why $\mathcal K_g=\Upsilon_{-,g}/T$ is an explicit assumption in the IC comparison rather than a general identity.

Identifying ${\rm c/r}$ with $\pm$, we form $\Upsilon_{\pm,g}=\sqrt{\Upsilon_{\pm,x}\Upsilon_{\pm,y}}$. The C crossing, defined by $\Upsilon_{+,g}=T/(2\pi)$, estimates the C--L boundary. The IC crossing is the downward crossing of $\Upsilon_{+,g}+\Upsilon_{-,g}=2T/\pi$ between $T=0.90$ and $0.92$ and serves as a finite-size melting diagnostic with the conditional RG interpretation given above. Both crossing temperatures are denoted $T_\Upsilon$.

\begin{figure}[t]
\centering
\begin{minipage}[t]{0.32\linewidth}
\centering
\includegraphics[width=\linewidth]{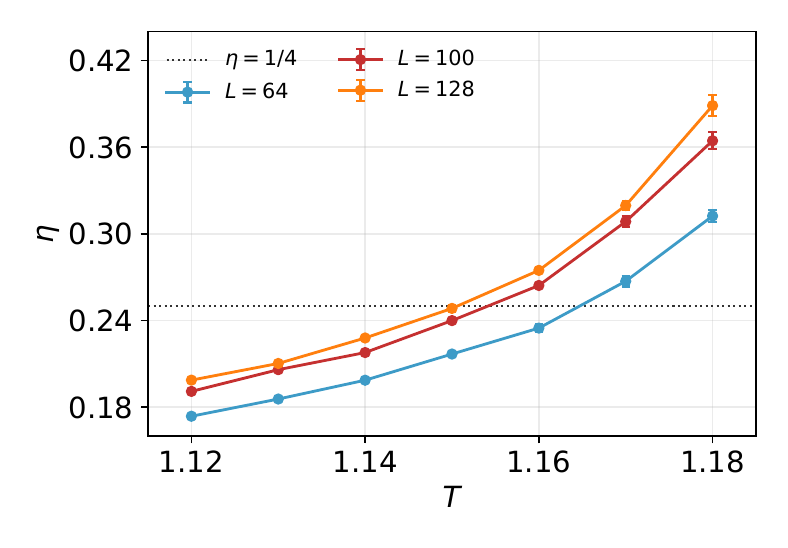}
\par\small (a)
\end{minipage}\hfill
\begin{minipage}[t]{0.32\linewidth}
\centering
\includegraphics[width=\linewidth]{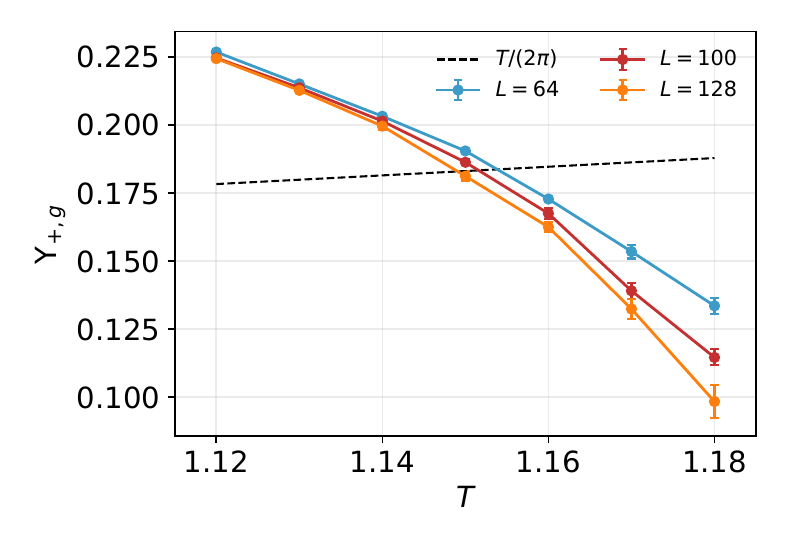}
\par\small (b)
\end{minipage}\hfill
\begin{minipage}[t]{0.32\linewidth}
\centering
\includegraphics[width=\linewidth]{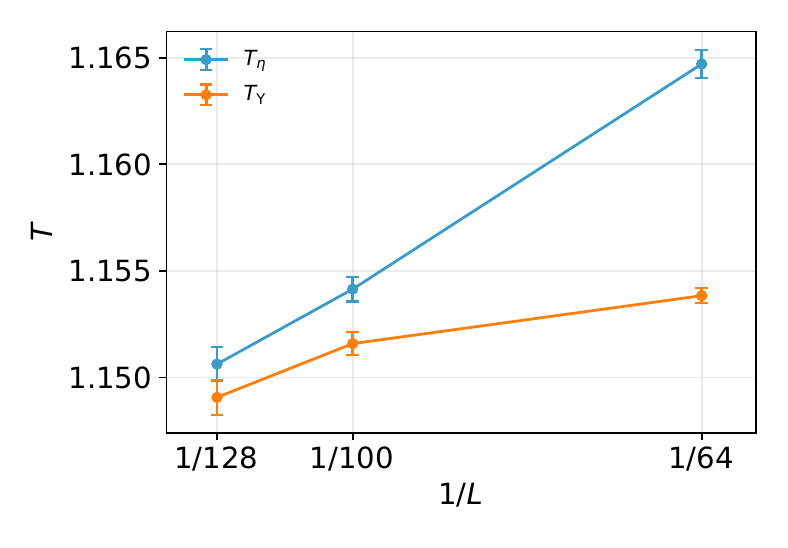}
\par\small (c)
\end{minipage}
\caption{Finite-size C--L diagnostics at $K=0.2$ and $k_0=0.25$ for $L=64$, 100, and 128. (a) Mean diagonal layer exponent $\eta_l$; the dotted line marks $1/4$. (b) Common helicity modulus $\Upsilon_{+,g}$; the dashed line marks $T/(2\pi)$. (c) Crossing temperatures $T_\eta$ from $\eta_l=1/4$ and $T_\Upsilon$ from $\Upsilon_{+,g}=T/(2\pi)$, plotted against $1/L$.}
\label{fig:sm-CL-fss}
\end{figure}

With $z_l(x,y)=\exp[i\theta_l(x,y)]$, the full layer-basis correlation matrix is
\begin{equation}
C_{ab}(r_y)=
\left\langle z_a(x,y+r_y)z_b^*(x,y)\right\rangle_{x,y},
\quad a,b=1,2,
\label{eq:sm-Cab}
\end{equation}
averaged over all $y$ origins with periodic wraparound and over all $x$ for Fig.~\ref{fig:eta}(a) and Fig.~\ref{fig:sm-CL-fss}, but over $20\le x<80$ for Fig.~\ref{fig:eta}(b). Each of the four channels $C_{11}$, $C_{22}$, $C_{12}$, and $C_{21}$ is fitted separately over a linear window on log--log axes. All reported exponents use $y$-direction correlations; in a directional check, most diagonal-exponent differences between $x$- and $y$-direction fits lie within two paired-chain bootstrap standard deviations, with isolated differences of $2$--$3$ standard deviations.

In the algebraic regime we fit
\begin{equation}
|C_{ab}(r_y)|=A_{ab}r_y^{-\eta_{ab}},
\label{eq:sm-eta-fit}
\end{equation}
using windows that begin at $r_y\ge4$, contain at least ten consecutive separations, and end before the finite-size tail; among windows with near-minimal log--log residual, we select the longest. A semilog exponential fit over the same interval serves as a diagnostic. Statistical uncertainties come from 600 paired whole-chain bootstrap resamples at fixed $(L,K,k_0,T)$; correlation and stiffness crossings share chain indices, and fits and crossing interpolations are repeated in each resample. Agreement within the diagonal and off-diagonal channel pairs is used as an equilibration check rather than imposed in the analysis.

\begin{samepage}
To test whether the correlation and helicity-modulus diagnostics locate the same C--L boundary, we compare their finite-size crossings on the cut $K=0.2$, $k_0=0.25$, where the equilibrium soliton sector has $N_{\rm sol}=0$. We locate $T_\eta$ from $\eta_l=1/4$, with $\eta_l=(\eta_{11}+\eta_{22})/2$ the mean diagonal layer exponent, and $T_\Upsilon$ independently from $\Upsilon_{+,g}=T/(2\pi)$. Figure~\ref{fig:sm-CL-fss}(a) shows that $\eta_l$ increases smoothly through $1/4$, and Fig.~\ref{fig:sm-CL-fss}(b) shows the corresponding common-helicity-modulus crossings. Figure~\ref{fig:sm-CL-fss}(c) collects the two crossing temperatures: their separation is smaller at $L=100$ and 128 than at $L=64$, and at $L=128$ the estimates differ by $0.00156$ near $T\simeq1.150$. With 16 chains at $L=100$, we obtain $T_\eta=1.15414(54)$ and $T_\Upsilon=1.15159(49)$, with one-standard-deviation statistical uncertainties in parentheses. At $L=128$, changing the correlation-fit window from 5--25 to 5--16 shifts $T_\eta$ by $-0.0026$, which places the residual difference within the fitting-window sensitivity. The two criteria thus locate the same narrow temperature region, although their finite-size estimates need not coincide within statistical uncertainties. This comparison concerns the C side; the IC results are reported as helicity-modulus and correlation diagnostics under the sampling and stiffness assumptions stated above.
\end{samepage}

\subsection{Fixed-sector free energies and PT scaling}

In the dilute-soliton regime, both the soliton density and the free-energy splitting between neighboring sectors are small, as expected in the PT wall theory~\cite{PokrovskyTalapov1979,Coppersmith1981,Horovitz1983}. Local Metropolis updates at fixed $k_0$ rarely change the soliton number, so the sampled state can depend strongly on the initial sector over accessible simulation times. We therefore compute the free-energy difference between sectors with $N$ and $N+1$ solitons by thermodynamic integration.

The fixed-sector crossings are computed on $L_x\times L_y$ cylinders, open along $x$ and periodic along $y$. Rather than relying on local Metropolis dynamics to nucleate complete soliton walls, we impose boundary phases that set the relative winding along $x$. For a sector coordinate $\lambda$, we write
\begin{equation}
\theta_1=u_1+\frac{\pi\lambda x}{L_x-1},\qquad \theta_2=u_2-\frac{\pi\lambda x}{L_x-1},
\label{eq:sm-lambda}
\end{equation}
where the compact fields $u_1,u_2$ are updated in the interior and fixed at the two $x$ boundaries, and we substitute this decomposition into all terms of Eq.~\eqref{eq:sm-lattice-H}, including the Josephson term. The background then advances the relative phase by $2\pi\lambda$: integer $\lambda=N$ labels the intended $N$-soliton sector, and noninteger $\lambda$ serves only as a thermodynamic-integration coordinate between neighboring sectors $N$ and $N+1$.

No explicit winding constraint is imposed: the interior phases are compact, bulk vortices are allowed, and the same energy-based Metropolis rule is used at every $\lambda$. Because the boundary values alone do not prevent phase slips, retention of the intended winding relies on initialization and metastability over the run. We monitor the row-resolved relative winding $n_r(y)=(2\pi)^{-1}\sum_x[\mathrm{wrap}(\Delta_x\theta_1)-\mathrm{wrap}(\Delta_x\theta_2)]$, where $\Delta_x\theta_l$ is the bond phase difference along $x$, wrapped into $(-\pi,\pi]$. In the saved records, no majority of rows departs from the target branch for 20 of the 39 crossings in Fig.~\ref{fig:ci}, time-resolved for 10 and from final states for 10; spatial records are not available for the remaining 19.

Let $f_\lambda=F_\lambda/L_y$ be the free energy per unit wall length. For a consistently continued sector ensemble, with the thermodynamic derivative $\partial_\lambda F_\lambda=\langle\partial_\lambda\mathcal H\rangle$, the neighboring-sector difference is
\begin{equation}
\Delta f_N(k_0)=f_{N+1}(k_0)-f_N(k_0)=\int_N^{N+1}d\lambda\,\left\langle\frac{1}{L_y}\frac{\partial \mathcal H}{\partial\lambda}\right\rangle.
\label{eq:sm-lambda-TI}
\end{equation}
For the 39 crossings in Fig.~\ref{fig:ci}, 13 at each temperature, we evaluate this integral by composite trapezoid quadrature on nine equally spaced $\lambda$ nodes ($\Delta\lambda=1/8$), refined to 17 nodes ($\Delta\lambda=1/16$) for the four curves $(T,L,N)=(0.45,100,4)$, $(0.45,128,6)$, $(0.55,100,4)$, and $(0.55,128,6)$. Forward and backward continuations are both retained to monitor hysteresis.

To improve precision, we evaluate the full $\lambda$ integral at an anchor field $k_a$ and continue it in $k_0$ with
\begin{equation}
\Delta f_N(k_0)=\Delta f_N(k_a)+\frac{1}{L_y}\int_{k_a}^{k_0}dk\,\big[\langle\partial_k\mathcal H\rangle_{N+1}-\langle\partial_k\mathcal H\rangle_N\big],
\label{eq:sm-k-TI}
\end{equation}
where $\Delta f_N(k_a)$ comes from the full $\lambda$ integration. After coarse fixed-sector scans, we select sector pairs $(N,N+1)$ whose midpoint densities, defined below, satisfy $0.002\leq\rho_{N+1/2}^{2}\leq0.008$, and use their estimated crossings to set local $k_0$ grids. Each pair has its own anchor; the anchors lie between $k_a=0.73$ and $0.805$, and the grids contain 2--10 field values with spacing $0.005$ or $0.01$. For six pairs at $T=0.45$ and three at $T=0.55$, full $\lambda$ integrals exist at two anchors; the lower anchor gives the main estimate, and using the upper one instead shifts $k_0^c$ by $+0.0022$ at $T=0.45$ and $-0.0008$ at $T=0.55$. We integrate the measured field derivatives from $k_a$ by cumulative trapezoid quadrature and locate each crossing $\Delta f_N(k_N)=0$, at which sectors $N$ and $N+1$ exchange stability, by linear interpolation between adjacent grid points of opposite sign, without extrapolation. Each crossing is assigned the midpoint density $\rho_{N+1/2}=(N+1/2)/(L_x-1)$, the finite-size estimator of $\rhosol$.

The finite-temperature PT law $\rhosol\propto[k_0-k_0^c(T)]^{1/2}$ can thus be tested without fitting an integer staircase, through the leading-order form
\begin{equation}
k_N\simeq k_0^c(T)+A(T)\rho_{N+1/2}^2,
\label{eq:sm-PT-law}
\end{equation}
which neglects higher-density and finite-size corrections. The fit in Fig.~\ref{fig:ci} combines the $L=64$, 100, and 128 crossings over $0.002\le\rho_{N+1/2}^2\le0.008$. Here $k_0^c(T)$ is the extrapolated zero-density intercept of this fit; because opposite-vorticity pairs are relevant in the dilute limit [Eq.~\eqref{eq:sm-PT-stiffness}], it is not an independently established equilibrium boundary.

Statistical errors are propagated through both integrations and the $k_0^c(T)$ fit with 600 bootstrap resamples. Each resample draws whole histories or chains within each initialization family, i.e., runs started from the same initial state, and takes the $\lambda$ and field integrands jointly from the same draws; the one single-history curve, $(T,L,N)=(0.55,128,8)$, is instead resampled by a circular block bootstrap over consecutive pairs of 8000-sweep blocks. A resample enters the fit only if all 13 fitted points have a unique crossing, which holds for 561, 593, and 600 of the 600 resamples at $T=0.35$, 0.45, and 0.55; the quoted intervals are conditional on these resamples. Table~\ref{tab:kc-sensitivity} reports the integration, fit-range, and scan-direction sensitivities separately.

\begin{table}[tb]
\centering
\caption{Sensitivity of the fitted intercepts $k_0^c(T)$. The column $\Delta k_0^{c,{\rm int}}$ replaces trapezoid by Simpson quadrature for the $\lambda$ integral in Eq.~\eqref{eq:sm-lambda-TI}, with the field integration in Eq.~\eqref{eq:sm-k-TI} unchanged; $\Delta k_0^{c,{\rm range}}$ restricts the fit from 13 points to the seven with $\rho_{N+1/2}^2\leq0.0045$; and the direction span is the absolute difference between the mean fits for the two scan directions, i.e., between the forward-only and backward-only intercepts. These are observed sensitivities, not total confidence intervals.}
\label{tab:kc-sensitivity}
\begin{tabular}{rrrr}
\toprule
$T$ & $\Delta k_0^{c,{\rm int}}$ & $\Delta k_0^{c,{\rm range}}$ & direction span\\
\midrule
0.35 & $+0.0083$ & $+0.0024$ & 0.0113\\
0.45 & $+0.0033$ & $+0.0089$ & 0.0029\\
0.55 & $+0.0040$ & $+0.0080$ & 0.0028\\
\bottomrule
\end{tabular}
\end{table}

\section{Zero-temperature sector minimization}
\label{app:zeroT}

As a separate check of the zero-temperature C--IC threshold, we minimize the lattice phase energy over the compact phase variables in fixed soliton-number sectors, without Monte Carlo sampling. We use periodic boundaries along $x$ at zero common winding, so adjacent allowed sectors contain $N$ and $N+2$ individual solitons, with even $N$. Their crossing field is denoted $k_N$, and their midpoint density is $\rho_{N+1}=(N+1)/L_x$. The low-density crossings converge to a lattice threshold within 0.5\% of the continuum value $\kc=(4/\pi)\sqrt{2K/\rho_s}$.

At zero temperature, the dilute soliton train has sine-Gordon asymptotics: the soliton spacing diverges logarithmically near threshold, so the density follows the inverse-log form
\begin{equation}
\rhosol(k_0)\simeq\frac{1}{\xi_{\rm sol}\ln[A_0/(k_0-\kc)]},\qquad k_0\rightarrow k_0^{c+},
\label{eq:sm-zeroT-inverse-log}
\end{equation}
where $A_0$ and $\xi_{\rm sol}$ are nonuniversal constants. For the fixed-sector crossings, we use the leading exponential form, neglecting possible density-dependent prefactors,
\begin{equation}
k_N-\kc \simeq B_0\exp\!\left[-\frac{1}{\xi_{\rm sol}\rho_{N+1}}\right],
\label{eq:sm-zeroT-crossing}
\end{equation}
so the crossings are tested by plotting $\ln(k_N-\kc)$ against $1/\rho_{N+1}$ rather than through the finite-temperature relation $k_N-k_0^c(T)\propto\rho_{N+1/2}^2$.

\begin{figure}[t]
\centering
\includegraphics[width=\linewidth]{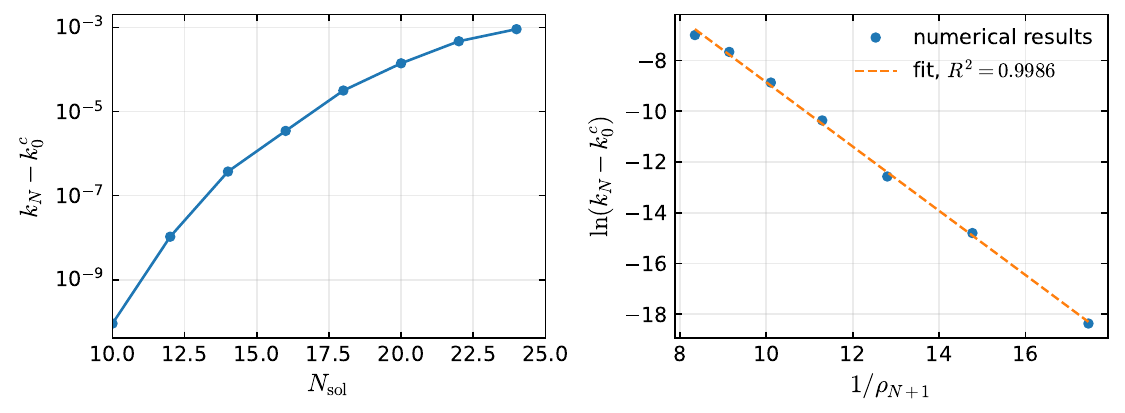}
\caption{Zero-temperature fixed-sector minimization at $K=0.8$ and $L=192$. Left: crossing offsets $k_N-\kc$, with each crossing between sectors $N$ and $N+2$ plotted at the upper soliton number $N_{\rm sol}=N+2$. Right: $\ln(k_N-\kc)$ against $1/\rho_{N+1}$, with the inverse-log fit of Eq.~\eqref{eq:sm-zeroT-crossing} ($R^2=0.9986$). The crossings approach the $T=0$ C--IC threshold as expected for a dilute soliton lattice. This check is independent of the finite-temperature free-energy analysis in the main text.}
\label{fig:sm-zeroT}
\end{figure}

For the $K=0.8$, $L=192$ data in Fig.~\ref{fig:sm-zeroT}, the mean of the five dilute crossings $0\leftrightarrow2,\ldots,8\leftrightarrow10$ sets the reference $\kc=1.603996$. Their offsets from this reference lie below numerical resolution and cannot reliably enter the logarithmic fit, so we hold the unrounded reference fixed and fit the seven crossings $10\leftrightarrow12,\ldots,22\leftrightarrow24$, obtaining $R^2=0.9986$. Size and tolerance checks vary the dilute crossings by less than $6\times10^{-9}$ within the $y$-uniform minimization, and joint inverse-log fits with different fit ranges shift $\kc$ by up to $3.2\times10^{-5}$, a fit-range sensitivity rather than a statistical error.

	\bibliographystyle{apsrev4-2}
	\bibliography{reference}
	
\end{document}